\documentclass[aps,pre,reprint,groupedaddress,showpacs]{revtex4-1}
\usepackage[usenames,dvipsnames]{color}
\usepackage{enumerate}
\usepackage{graphicx}
\usepackage{multirow}
\usepackage{cases}

\DeclareGraphicsExtensions{.png}

\newcommand{\secref}[1]{Sec.~\ref{sec_#1}}
\newcommand{\figref}[1]{Fig.~\ref{fig_#1}}
\newcommand{\figrefs}[2]{Fig.~\ref{fig_#1}~(#2)}
\newcommand{\tblref}[1]{Table~\ref{tbl_#1}}
\newcommand{\eqref}[1]{Eq.~(\ref{eq_#1})}
\newcommand{\appref}[1]{App.~\ref{app_#1}}

\newcommand{\ff}[0]{FF}
\newcommand{\sff}[0]{SFF}
\newcommand{\lff}[0]{LFF}

\newcommand{\cpa}[0]{CP}
\newcommand{\dpa}[0]{DP}
\newcommand{\fpa}[0]{FP}
\newcommand{\gpa}[0]{GP}
\newcommand{\hpa}[0]{HP}
\newcommand{\mpa}[0]{MP}
\newcommand{\lpa}[0]{LP}
\newcommand{\gmo}[0]{GM}
\newcommand{\luv}[0]{LUV}
\newcommand{\scp}[0]{SCP}
\newcommand{\mcl}[0]{MCL}
\newcommand{\imd}[0]{IMD}
\newcommand{\imp}[0]{IMP}
\newcommand{\nmf}[0]{NF}
\newcommand{\km}[0]{KM}
\newcommand{\mm}[0]{MM}
\newcommand{\cm}[0]{CM}

\newcommand{\GN}[0]{GN}
\newcommand{\GNT}[0]{GN2}

\newcommand{\HNV}[0]{HN7}
\newcommand{\HNX}[0]{HN6}
\newcommand{\LFR}[0]{LFR}
\newcommand{\cora}[0]{\textit{Cora} citation network}
\newcommand{\coras}[0]{\textit{Cora} network}

\newcommand{\nmi}[0]{NMI}

\newcommand{\ari}[0]{ARI}

\newcommand{\g}[0]{\mathcal{G}}
\newcommand{\h}[0]{\mathcal{H}}
\newcommand{\I}[0]{\mathcal{I}}

\newcommand{\like}[0]{\mathcal{L}}
\newcommand{\logL}[0]{$\log\mathcal{L}$}

\newcommand{\mlogL}[0]{$-\log\mathcal{L}$}

\newcommand{\neigh}[0]{\Gamma}
\newcommand{\argmax}[0]{\arg\!\max}

\definecolor{black}{RGB}{0, 0, 0}
\definecolor{gray}{RGB}{121, 121, 121}
\definecolor{blue}{RGB}{0, 84, 147}
\definecolor{mocha}{RGB}{147, 82, 0}
\definecolor{asparagus}{RGB}{146, 144, 0}

\begin{document}


\title{Clustering assortativity, communities and functional modules in real-world networks}

\author{Lovro \v Subelj}
\thanks{Electronic address: lovro.subelj@fri.uni-lj.si}
\author{Marko Bajec}
\thanks{Electronic address: marko.bajec@fri.uni-lj.si}
\affiliation{University of Ljubljana, Faculty of Computer and Information Science, Ljubljana, Slovenia}

\date{\today}

\begin{abstract}
Complex networks of real-world systems are believed to be controlled by common phenomena, producing structures far from regular or random. Clustering, community structure and assortative mixing by degree are perhaps among most prominent examples of the latter. Although generally accepted for social networks, these properties only partially explain the structure of other networks. We first show that degree-corrected clustering is in contrast to standard definition highly assortative. Yet interesting on its own, we further note that non-social networks contain connected regions with very low clustering. Hence, the structure of real-world networks is beyond communities. We here investigate the concept of functional modules---groups of regularly equivalent nodes---and show that such structures could explain for the properties observed in non-social networks. Real-world networks might be composed of functional modules that are overlaid by communities. We support the latter by proposing a simple network model that generates scale-free small-world networks with tunable clustering and degree mixing. Model has a natural interpretation in many real-world networks, while it also gives insights into an adequate community extraction framework. We also present an algorithm for detection of arbitrary structural modules without any prior knowledge. Algorithm is shown to be superior to state-of-the-art, while application to real-world networks reveals well supported composites of different structural modules that are consistent with the underlying systems. Clear functional modules are identified in all types of networks including social. Our findings thus expose functional modules as another key ingredient of complex real-world networks.
\end{abstract}



\maketitle


\footnotetext[1]{Some of the earlier work refers to functional and structural modules differently~\cite{RW07,SB12u}.}

\footnotetext[2]{Definition of $r$ is consistent with directed networks, while for more practical expression see~\cite{New02}. Note that Pearson coefficient detects only linear correlations. Although one can study neighbor connectivity plots~\cite{PVV01}, correlation profiles~\cite{MS02} or other measures~\cite{LADW05} instead, $r$ is often preferable due to its convenience.}
\footnotetext[3]{$p_c$~is derived for alternative, but similar, definition of~$C$ that allows easier analytical consideration~\cite{NSW01,NP03}.}
\footnotetext[4]{$r_d$ of a random graph~\cite{ER59} does not go to zero in the limit of large $n$ (results omitted). However, since $D\approx 0$, $r_d$~captures only small random fluctuations in $d$.}
\footnotetext[5]{No closed formula for $W$ in a random graph exists~\cite{ZLZ11b}. Thus, we maximize $W$ in $100$ realizations of a corresponding random graph~\cite{ER59} and take the $95$-th percentile as the expected value. Hence, the results are statistically significant at $p\mbox{-value}=0.05$.}

\footnotetext[6]{Results reported under \ff~model correspond to the original model for directed networks~\cite{LKF07,LKF05}.}
\footnotetext[7]{In practice, $x_p$ and $x_q$ are sampled from negative binomial distributions $\mathrm{NB}(1,p)$ and $\mathrm{NB}(1,q)$.}
\footnotetext[8]{Multiple $p$ and $q$ will give a network with same~$C$ and $r$. Still, for fixed $k$, there is a unique solution (\eqref{k}).}

\footnotetext[9]{To prevent oscillations of labels, $i$ retains its current label when it is among most frequent in $\neigh_i$~\cite{RAK07}.}
\footnotetext[10]{Stability parameter $\beta$ is set to $2$, and to $0.25$ for larger real-world networks~\cite{SB11b} (to ensure convergence).}


\section{\label{sec_intro}Introduction}
Networks are the simplest representation of complex systems of interacting parts. Examples of these are ubiquitous in practice, including social networks~\cite{FIA11}, information systems~\cite{SB11s}, cooperate ownerships~\cite{VGB11} and food webs~\cite{WM00}, to name just a few. Despite seemingly plain form, real-world networks commonly exhibit complex structural properties that are absent from regular or random systems~\cite{WS98,FFF99}. Network complexity arises not from that of individual interactions, but rather from their intrinsic collective behavior. Thus, networked systems are believed to be controlled by common phenomena, which has been the main focus of network science in the last decade~\cite{WS98,BA99,GN02,New02}. Nevertheless, our comprehension of real-world network structure remains to be only partial~\cite{New08}.

Network transitivity or clustering~\cite{WS98,NSW01}, degree mixing~\cite{New02,New03b}---degree correlations of links' ends---and community structure~\cite{FLG00,GN02} are perhaps among most widely analyzed network properties in physics literature~\cite{New10,New03a}. Communities are usually seen as densely linked groups of nodes that are only sparsely linked with the rest of the network~\cite{GN02,RCCLP04}. These are, at least in context of social networks, considered an artifact of triadic closure~\cite{Gra73} or homophily~\cite{MSC01,NG03}, whereas communities also imply assortative---positively correlated---mixing by degree, as long as their sizes differ~\cite{NP03}. On the other hand, recent work suggests that network transitivity, rather than homophily, is the cause of community structure and degree assortativity in real-world networks~\cite{FFGP11}. Regardless of the latter, there is substantial evidence that communities and assortative mixing appear concurrently with high clustering, properties also captured by many network models in the literature~\cite{New03b,LKF07,WH09,RCS11}.

\begin{figure*}[t]
\includegraphics[width=1.66\columnwidth]{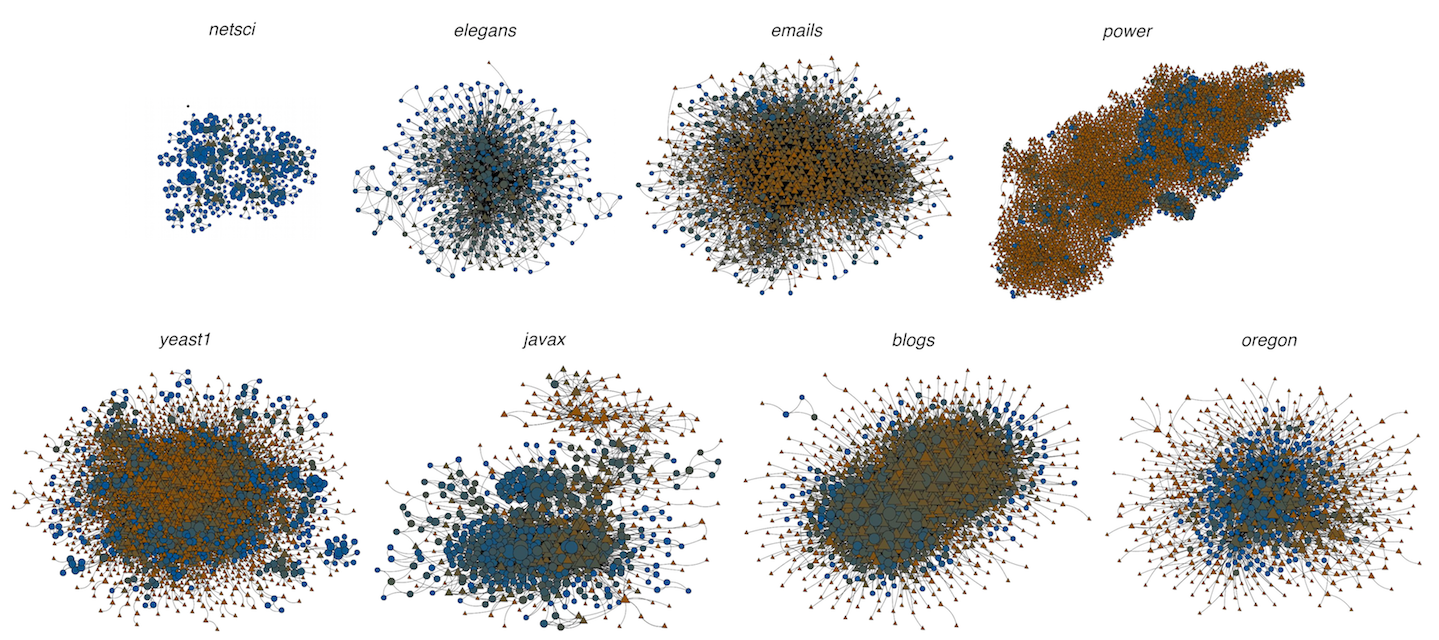}
\caption{\label{fig_rw}(Color online)~Largest connected components of different real-world networks in~\tblref{rw}. Node sizes are proportional to degrees, whereas symbols correspond to degree-corrected clustering~$d$ that ranges from zero ({\color{mocha}polygons}) to one ({\color{blue}circles}). Observe that $d$ is highly assortative, whereas nodes with particularly high or low $d$ are localized in regions that are characteristic within the underlying system. For example, ignoring nodes with degree one, printing classes in \textit{javax} software network experience average clustering of only $D=0.176$, while $D=0.602$ for visualization classes (above and below, respectively). Similarly, liberal and conservative blogs in \textit{blogs} web graph experience $D=0.435$ and $D=0.373$ (left-hand and right-hand side, respectively).} 
\end{figure*}

However, non-social networks greatly deviate from this picture. Biological and technological networks are in fact degree disassortative---negatively correlated---whereas information networks usually exhibit no clear degree mixing~\cite{New02,HL11}. Moreover, many real-world networks contain connected regions of nodes with very low clustering, where classical definition of community does not apply (see~\figref{rw}). Although one can still partition the network into well separated groups~\cite{LKSF10}, several authors have argued that clear communities emerge only in some (parts of) real-world networks~\cite{LRR10,NM11b,ZLZ11b,SB12u}. This poses an interesting question: ``Are there mesoscopic structures beyond classical communities that could explain for the properties observed in non-social networks?''.

The main purpose of this paper is to expose functional modules~\cite{NL07,RW07,ABFX08,SB12u}---groups of regularly equivalent~\cite{EB94} nodes---as a possible answer. Nodes are regularly equivalent if they are linked in the same way to other equivalent nodes (e.g., multi-partite structures). Regular equivalence is a relaxed version of structural equivalence~\cite{LW71} that demands for the nodes to be linked exactly the same. Hence, functional modules refer to groups of nodes that are linked similarly with the rest of the network---have common neighborhoods---and thus perform the same function within the underlying system~\cite{PSR10,SB12u}. Both functional modules and communities can be considered under a general concept of structural modules~\footnotemark[1]---groups of nodes with common linking patterns---although some asymmetries exist~\cite{SB12u} (e.g., mutual independence). (Note that communities can also be seen as functional modules~\cite{NL07,RW07}. However, most pronounced functional modules are disconnected groups of nodes, whereas best communities are, obviously, connected.)

Structure of the paper is nonstandard. We first analyze degree and clustering mixing in a large number of real-world networks of different type and origin (\secref{mix}). Analysis reveals that degree-corrected clustering~\cite{SV05} is in contrast to standard definition highly assortative in all networks (see~\figref{rw})---a distinctive property of real-world networks that was previously unobserved (to our knowledge). We further show that most non-social networks contain a large number of nodes with clustering lower than expected at random, which, together with clustering assortativity, implies connected regions of nodes with extremely low transitivity. Common intuition of dense communities does not coincide with the latter, while adequate extraction of communities from these networks even increases their degree disassortativity.

Functional modules should result in lower clustering and also degree disassortativity. Thus, we propose a simple network model that implicitly introduces functional modules into the network through link copying mechanism~\cite{KKRRT99,KRRSTU00} that resembles triadic formation~\cite{HK02,KOSKK07} (\secref{mod}). Each node explores the network using the burning process of forest fire model~\cite{LKF07,LKF05}, whereas links of visited nodes are copied independently of the latter. This process has a natural interpretation in many information, technological and other networks. Model indeed generates scale-free small-world networks~\cite{BA99,WS98} with community structure and degree (dis)assortativity, where clustering and degree mixing are controlled through parameters. This solves an open problem in network science---what simple local process produces degree disassortativity in real-world networks (to our knowledge). We further show that clustering assortativity is related to the extent of overlap between communities and functional modules. Higher values correspond to clearer separation between different structural modules, which appears to be realized through localization of communities. 

The above introduces 'structural-world' conjecture: ``Real-world networks are composed of functional modules characterizing different roles within the underlying system, and locally overlaid by communities based on some assortative property of the nodes.''. The former explain degree disassortativity and efficient long-range navigation---strength of weak ties~\cite{Gra73}---whereas the latter increase the overall clustering and degree assortativity, and provide for efficient local navigation---weakness of strong ties~\cite{Gra73}. Note that conjecture deviates from classical comprehension of small-world phenomena~\cite{WS98}.

We proceed by presenting an algorithm for detection of arbitrary structural modules without any prior knowledge like the number of modules (\secref{alg}). Algorithm exploits label propagation~\cite{RAK07,SB12u} to partition the network into modules, whereas each module is further refined independently of others. Algorithm is shown to be comparable to state-of-the-art in community detection, and superior in detection of structural modules (\appref{soa}).

\secref{exp} first validates the algorithm on various synthetic networks with planted partition and random graphs. Next, application to real-world networks reveals well supported composites of different structural modules that are consistent with characteristics of the underlying system, and structural-world conjecture. Although (almost) all networks contain communities, the network structure is more accurately predicted by considering also other structural modules. In the case of functional modules, most apparent examples are found in technological, biological, and, surprisingly, also classical social networks.

We further use techniques presented in the paper to also conduct an exploratory analysis of a larger information network (\secref{cora}). We extract a citation network from \textit{Cora} dataset~\cite{MNRS00} that includes computer science publications collected from the web. Analysis reveals skewed size distribution only in the case of functional modules, which is inconsistent with some earlier work~\cite{CNM04,PDFV05}. Most pronounced functional modules else mainly arrange in bipartite structures, however, the complexity of linking patterns is much higher than expected.

We conclude the paper in~\secref{conc}.


\section{\label{sec_mix}Degree and clustering mixing}
Let a network be represented by a simple undirected graph, with $N=\{1,\dots,n\}$ being the set of itrs nodes and $L$ being the set of its links (denote $m=|L|$). Also, let $k_i$ be degree of node $i$ and $k$ the average degree. 

\tblref{rw} shows common statistic for $24$ real-world networks of different size and origin. We consider most types of networks usually found in the literature, whereas detailed description is omitted here. Networks are ordered with respect to degree mixing coefficient~$r$~\cite{New02}. $r$~measures degree correlations in a network and is just a Pearson correlation coefficient of degrees at links' ends.
\begin{eqnarray}
r & = & \frac{1}{2m\sigma_k}\sum_{ij}\left(k_i-k\right)\left(k_j-k\right),
\label{eq_r}
\end{eqnarray}
where $\sigma_k$ is standard deviation and the sum goes through all linked pairs $(i,j)$, $r\in[-1,1]$~\footnotemark[2]. Positive correlation is indicated by $r\gg 0$, which is known as assortative mixing by degree. Similarly, disassortative mixing refers to negative correlation or, equivalently, $r\ll 0$.

In scale-free networks, which most real-world networks are, $r$ can be seen as a tendency of hubs~\cite{HBHGBZDWCRV04}---high degree nodes---to link between themselves. Observing values of $r$ in~\tblref{rw} one can conclude that the latter is indeed not the case, as most real-world networks are degree disassortative. Only social networks show strong assortativity, whereas most information and some technological networks exhibit no clear degree mixing with $r\approx 0$.

We proceed with an introduction of node clustering coefficient $c$~\cite{WS98}. $c_i$ measures transitivity around $i$ and is defined as the fraction of linked neighbors.
\begin{eqnarray}
c_i & = & \frac{t_i}{{k_i \choose 2}},
\label{eq_c}
\end{eqnarray}
where $t_i$ in the number of links among neighbors of $i$---the number of closed triads---and ${k_i \choose 2}$ is the number of all possible links, $c_i\in[0,1]$. For $k_i\leq 1$, $c_i=0$ by definition. Transitivity of the entire network can be estimated by simply averaging over all the nodes, which is known as (network) clustering coefficient $C$~\cite{WS98}, $C\in[0,1]$.

Real-world networks are characterized by much higher $C$ than expected by chance (see~\tblref{rw}). For example, random graph \`{a}~la Erd\"{o}s-R\'{e}nyi~\cite{ER59}, where links are laid between nodes with probability $p_r=k/(n-1)$, exhibits only $C=p_r$ in the limit of large $n$. This is less than $0.02$ for almost all networks considered here. Still, in the case of configuration model~\cite{MR95,NSW01}, where graphs are sampled from an ensemble with the same degree sequence as the network, $C\approx p_c$ for large enough $n$~\cite{EMB02,NP03}.
\begin{eqnarray}
p_c & = & \frac{\left(\sum_i k_i^2-nk\right)^2}{n^3k^3}
\label{eq_p_c}
\end{eqnarray}
Although $p_c$ scales as $n^{-1}$, it is not necessarily negligible for networks of moderate size~\footnotemark[3] (see below).

\figrefs{cd}{top} shows scaling of $c$ with respect to node degree in different real-world networks (\tblref{rw}). Note that $c$ decays with degree resembling a power-law form. Actually, different authors have observed that $c\sim k^{-\alpha}$ with $\alpha\in[0.75,1]$ in Internet, metabolic, collaboration and other networks~\cite{RSMOB02,VPV02,RB03}, whereas the same behavior also emerges in a hierarchical network~\cite{RB03}. However, \citet{SV05} have shown that this scaling is actually due to node degree mixing. This can be motivated by an observation that hubs would always have low clustering, as the opposite implies a very large clique. Notice, for example, that power-law decays are much steeper for degree disassortative networks than for assortative ones.

\begin{figure}[t]
\includegraphics[width=1.00\columnwidth]{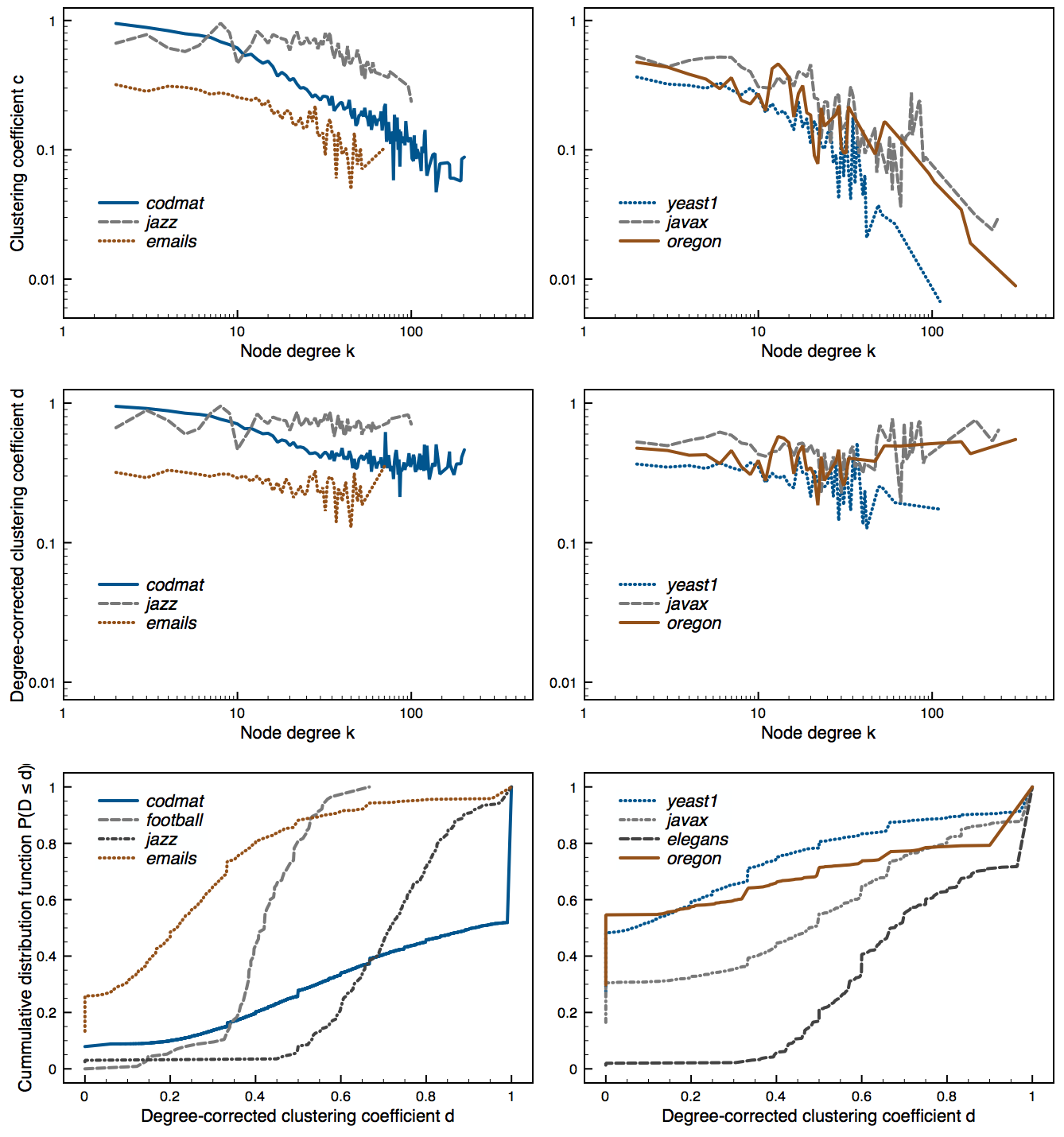}
\caption{\label{fig_cd}(Color online)~Scaling of clustering in degree assortative and disassortative real-world networks (left and right, respectively). (Nodes with degree at most one are ignored.)}
\end{figure}

\begin{table*}[t]
\begin{ruledtabular}
\begin{tabular}{cccccccccccrr}
Type & Network & Description & $n$ & $m$ & $k$ & $C$ & $D$ & $r$ & $r_c$ & $r_d$ & \multicolumn{1}{c}{$d<p_r$} & \multicolumn{1}{c}{$d<p_c$} \\\hline
\multirow{3}{*}{Collaboration} & \textit{netsci} & Network scientists~\cite{New06b} & $1589$ & $2742$ & $3.5$ & $0.638$ & $0.690$ & $0.462$ & $0.442$ & $0.679$ & $1\%$ & $1\%$ \\

 & \textit{condmat} & \textit{Cond. Mat.} archive~\cite{New01d} & $27519$ & $116181$ & $8.4$ & $0.655$ & $0.722$ & $0.166$ & $0.116$ & $0.291$ & $1\%$ & $1\%$ \\

 & \textit{comsci} & Slovenian computer sci.~\cite{BSB12} & $239$ & $568$ & $4.8$ & $0.479$ & $0.561$ & $-0.044$ & $0.123$ & $0.355$ & $6\%$ & $6\%$ \\

Online social & \textit{pgp} & PGP web of trust~\cite{BPDA04} & $10680$ & $24316$ & $4.6$ & $0.266$ & $0.317$ & $0.238$ & $0.497$ & $0.632$ & $27\%$ & $27\%$ \\

\multirow{4}{*}{Social} & \textit{football} & American football~\cite{GN02} & $115$ & $613$ & $10.7$ & $0.403$ & $0.419$ & $0.162$ & $0.369$ & $0.385$ & $0\%$ & $0\%$ \\

 & \textit{jazz} & Jazz musicians~\cite{GD03} & $198$ & $2742$ & $27.7$ & $0.617$ & $0.703$ & $0.020$ & $0.008$ & $0.198$ & $1\%$ & $1\%$ \\

 & \textit{dolphins} & Bottlenose dolphins~\cite{LSBHSD03} & $62$ & $159$ & $5.1$ & $0.259$ & $0.319$ & $-0.044$ & $0.192$ & $0.234$ & $15\%$ & $15\%$ \\

 & \textit{karate} & Zachary's karate club~\cite{Zac77} & $34$ & $78$ & $4.6$ & $0.571$ & $0.666$ & $-0.476$ & $-0.229$ & $0.277$ & $3\%$ & $6\%$ \\

\multirow{2}{*}{Communication} & \textit{emails} & Emails at university~\cite{GDDGA03} & $1133$ & $5451$ & $9.6$ & $0.220$ & $0.253$ & $0.078$ & $0.214$ & $0.317$ & $14\%$ & $15\%$ \\

 & \textit{enron} & Emails at \textit{Enron}~\cite{LKF05} & $36692$ & $183831$ & $10.0$ & $0.497$ & $0.530$ & $-0.111$ & $0.185$ & $0.379$ & $4\%$ & $4\%$ \\

Road network & \textit{euro} & European highways~\cite{SB11b} & $1039$ & $1305$ & $2.5$ & $0.019$ & $0.025$ & $0.090$ & $0.395$ & $0.499$ & $91\%$ & $91\%$ \\

Power grid & \textit{power} & Western US grid~\cite{WS98} & $4941$ & $6594$ & $2.7$ & $0.080$ & $0.100$ & $0.003$ & $0.469$ & $0.653$ & $74\%$ & $74\%$ \\

Citation & \textit{hepart} & \textit{H.-E. Part.} archive~\cite{Aut03b} & $27770$ & $352285$ & $25.4$ & $0.312$ & $0.353$ & $-0.030$ & $0.132$ & $0.370$ & $6\%$ & $6\%$ \\

Documentation & \textit{javadoc} & Javadoc (\texttt{javax})~\cite{SB10k} & $2089$ & $7934$ & $7.6$ & $0.373$ & $0.433$ & $-0.070$ & $0.090$ & $0.440$ & $9\%$ & $9\%$ \\

\multirow{2}{*}{Protein} & \textit{yeast1} & Yeast \textit{S. cerevisiae}~\cite{PDFV05} & $2445$ & $6265$ & $5.1$ & $0.215$ & $0.250$ & $-0.101$ & $0.372$ & $0.534$ & $29\%$ & $29\%$ \\

 & \textit{yeast2} & Yeast \textit{S. cerevisiae}~\cite{JMBO01} & $2114$ & $2203$ & $2.1$ & $0.059$ & $0.072$ & $-0.162$ & $0.576$ & $0.675$ & $68\%$ & $68\%$ \\

\multirow{4}{*}{Software} & \textit{javax} & Java language (\texttt{javax})~\cite{SB11s} & $1595$ & $5287$ & $6.6$ & $0.381$ & $0.440$ & $-0.120$ & $-0.041$ & $0.545$ & $17\%$ & $17\%$ \\

 & \textit{jung} & JUNG graph library~\cite{SB11s} & $317$ & $719$ & $4.5$ & $0.366$ & $0.423$ & $-0.190$ & $0.092$ & $0.443$ & $21\%$ & $21\%$ \\

 & \textit{guava} & \textit{Guava} core libraries & $174$ & $355$ & $4.1$ & $0.320$ & $0.375$ & $-0.218$ & $0.075$ & $0.734$ & $34\%$ & $34\%$ \\

 & \textit{java} & Java language (\texttt{java})~\cite{SB11s} & $1516$ & $10049$ & $13.3$ & $0.685$ & $0.731$ & $-0.283$ & $-0.574$ & $0.536$ & $1\%$ & $100\%$ \\

Web graph & \textit{blogs} & Blogs on US politics~\cite{AG05} & $1490$ & $16715$ & $22.4$ & $0.263$ & $0.293$ & $-0.221$ & $-0.057$ & $0.308$ & $8\%$ & $13\%$ \\

Metabolic & \textit{elegans} & Nematode \textit{C. elegans}~\cite{JTAOB00} & $453$ & $2025$ & $8.9$ & $0.646$ & $0.710$ & $-0.226$ & $-0.240$ & $0.183$ & $1\%$ & $3\%$ \\

Internet & \textit{oregon} & Aut. systems (\textit{oregon})~\cite{LKF05} & $767$ & $1734$ & $4.5$ & $0.293$ & $0.317$ & $-0.299$ & $-0.231$ & $0.262$ & $35\%$ & $70\%$ \\

Bipartite & \textit{women} & Southern women club~\cite{DGG41} & $32$ & $89$ & $5.6$ & $0.000$ & $0.000$ & $-0.337$ & & & $100\%$ & $100\%$ \\\hline

\multirow{2}{*}{Random} && Erd\"{o}s-R\'{e}nyi graph~\cite{ER59} & & & & $p_r$ & $\geq p_r$ & $0$ & & \\

 & & Configuration model~\cite{MR95,NSW01} & & & & $p_c$ & $\geq p_c$ & $0$ \\

\end{tabular}
\end{ruledtabular}
\caption{\label{tbl_rw}Common statistics for different real-world networks gathered from the literature and random graph models. (Networks are treated as simple undirected graphs, while module detection algorithms consider multi graphs. Results for random models are valid in the limit of large $n$ (see text for details). Percentages in last two columns ignore nodes with degree at most one.)}
\end{table*}

Particularly, denominator in~\eqref{c} implicitly assumes that every two nodes can form a link between themselves. Although, this might be true in, e.g., online social networks, where links are generated for 'free', it indeed does not hold for other networks, where node degrees are subjected to, e.g., practical or technological constraints. Degree constraints thus introduce biases into~$c$ that are particularly apparent in degree disassortative networks.

Alternative definition of clustering that filters out degree biases has been proposed in the form of degree-corrected node clustering coefficient~$d$~\cite{SV05}.
\begin{eqnarray}
d_i & = & \frac{t_i}{\omega_i},
\label{eq_d}
\end{eqnarray}
where $\omega_i$ is the number of all possible links between neighbors of $i$ with respect to their degrees, $d_i\in[0,1]$. (Note that $\omega$ can be computed from neighbors degree sequence using a simple algorithm presented in~\cite{SV05}.) For $k_i\leq 1$, $d_i=0$ by definition. Again, clustering of the entire network can be estimated by simply averaging over all the nodes, which is denoted $D$~\cite{SV05}, $D\in[0,1]$.

Since $\omega\leq {k \choose 2}$, obviously, $d\geq c$ and $D\geq C$. The latter can be clearly observed in~\tblref{rw}. \figrefs{cd}{middle} also shows scaling of $d$ with respect to node degree. No characteristic form occurs, whereas $d$ is rather constant across several scales. Actually, under pseudofractal model introduced in~\cite{DGM02}, $c\sim 1/k$ implies $d\sim 1/\log k$~\cite{SV05}.

Next, \figref{rw} shows $d$ in different real-world networks in~\tblref{rw}. Interestingly, $d$ looks highly assortative in all networks, whereas nodes with similar $d$ are localized in regions that are characteristic within the underlying system. Note that although communities of highly clustered nodes can be clearly observed, remaining structure does not appear to be captured well by classical models.

Formally, we analyze clustering correlations in these networks by defining mixing coefficient $r_d$. As in~\eqref{r}, we adopt Pearson correlation coefficient to measure mixing of clustering at links' ends. (Similarly for $r_c$.)
\begin{eqnarray}
r_d & = & \frac{1}{2m\sigma_d}\sum_{ij}\left(d_i-D\right)\left(d_j-D\right),
\label{eq_r_d}
\end{eqnarray}
where $\sigma_d$ is standard deviation, $r_d\in [-1,1]$.

\begin{figure*}[t]
\includegraphics[width=2.00\columnwidth]{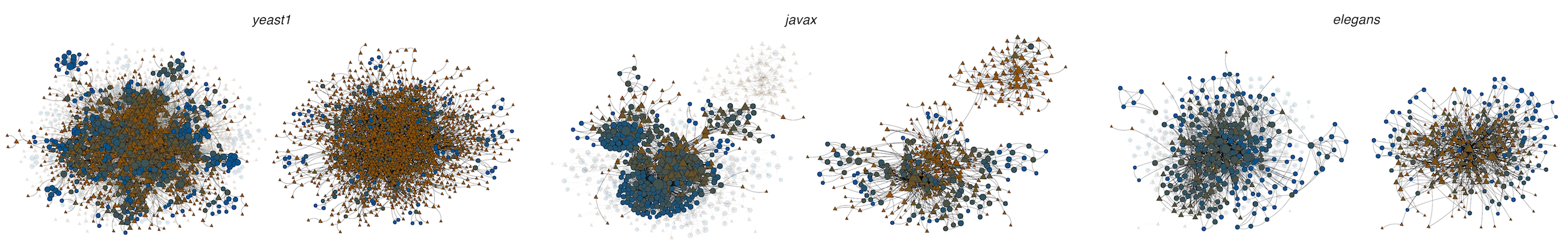}
\caption{\label{fig_over}(Color online)~Communities extracted from different real-world networks overlaid over originals, and networks after extraction (left-hand and right-hand side, respectively). Although communities span throughout the networks, many of the links remain unaccounted for. (The number of communities equals $13$, $14$ and $14$, while node symbols are consistent with~\figref{rw}.)}
\end{figure*}

Values of $r_d$ reported in~\tblref{rw} reveal that degree-corrected clustering $d$ is indeed highly assortative in all types of real-world networks. Although obvious to some extent, the latter cannot be considered an artifact of chance, since $r_d$ of a respective Erd\"{o}s-R\'{e}nyi random graph is significantly lower~\footnotemark[4] (due to no constraints on links). Moreover, same property is absent from standard definition of clustering $c$, whereas $r_c$ is even negative in some degree disassortative networks (see above). $r_d$ is also consistently much higher than $r_c$, where correlations above $0.5$ are note rare. Thus, in contrast to degree mixing, where $|r|<0.25$ in most cases, one can in fact accurately predict $d$ based on that of the neighbors. This gives an intuitive explanation of why there exist so many (local) approaches for community detection~\cite{For10,Sch07}.

Clustering assortativity or, equivalently, $r_d\gg 0$ can thus be regarded as another common property of real-world networks that distinguishes them from mere random world. The latter has not been previously observed and therefore offers various possibilities for future work. Actually, as shown in~\secref{mod}, synthetic generation of networks with $r_d\gg 0$, alongside with other common properties, is not straightforward.

\figrefs{cd}{bottom}~also plots distribution of $d$ in different degree assortative and disassortative networks. One can observe a relatively fast transition in the case of assortative networks, as most nodes share similar high value of~$d$ (e.g., \textit{football} and \textit{jazz} networks). This is an expected behavior under community structure model that characterizes social networks. However, in contrast to the latter, $30$-$55\%$ of nodes in disassortative networks have $d$ close to zero, whereas the distribution is else rather homogeneous (e.g., \textit{yeast1} and \textit{oregon} networks). Thus, despite some exceptions (e.g., \textit{elegans} network), structure of non-social networks goes beyond communities, which also casts doubts on the structure of social networks.

We further investigate peculiar clustering of non-social networks. \tblref{rw}~reports fractions of nodes with $d$ lower than predicted by random models discussed above---lower than $p_r$ and $p_c$. Although nodes with degree at most one, where $d$ equals zero by definition, are ignored, many networks still contain a large number of nodes with $d$ lower than expected by chance. Notice also a subtle difference between $p_r$ and $p_c$. For example, $p_r$ for \textit{oregon} Internet map is below $0.01$, whereas $p_c=0.87$. Thus, degree distributions alone can explain for the clustering observed in, e.g., scale-free networks~\cite{NP03}.

Note that existence of nodes with very low clustering is not particularly surprising by itself (see below). However, due to high clustering assortativity of real-world networks, the latter actually implies entire regions of nodes with very low clustering. These were mainly neglected in the past literature, as most authors were focused on nodes with high clustering, and hence community structure~\cite{POM09,For10}. Nevertheless, most networks analyzed here still contain some communities, which we address next.

\citet{ZLZ11b}~have already stressed the importance of absence of communities from certain (parts of) real-world networks. They have proposed community extraction framework that guards against the latter. Communities are extracted one by one, where each is selected from a candidate pool according to a quality measure~$W$ (see~\eqref{W}). When $W$ drops below the value one can expect under the same procedure in a corresponding random graph, the process terminates~\footnotemark[5]. Thus, communities are extracted only until statistically significant.

Quality of community $W$ is simply a difference between the links within the community and the links towards the rest of the network normalized appropriately~\cite{ZLZ11b}. Let $S$ be a community and $S^C$ its complement (denote $s=|S|$).
\begin{eqnarray}
W & = & s(n-s)\left(\frac{\sum_{i\in S}k_i^S}{s^2} - \frac{\sum_{i\in S}k_i-k_i^S}{s(n-s)}\right),
\label{eq_W}
\end{eqnarray}
where $k_i^S$ and $k_i-k_i^S$ are internal and external degree of node~$i$. ($k_i^S=|\neigh_i\cap S|$, where $\neigh_i$ is the set of neighbors of~$i$.) Factor $s(n-s)$ is an adjustment in the spirit of ratio cut~\cite{WC89,For10}. Since $s(n-s)$ is maximized at $s=n/2$, factor penalizes very small or large communities and thus produces more balanced cuts (see~\cite{ZLZ11b} for details).

\eqref{W}~can be rewritten into more convenient form as
\begin{eqnarray}
W & = & \sum_{i\in S}\frac{k_i^Sn}{s}-k_i.
\end{eqnarray}

Since work in~\cite{ZLZ11b} was focused on community structure alone, each time a community~$S$ was extracted, procedure was applied to its complement~$S^C$. However, only the links between the nodes in $S$ are accounted for, whereas those towards the rest of the network---between $S$ and $S^C$---should not be simply disregarded as random noise. (These links in fact constitute functional modules.)

Framework we adopt here therefore removes only the links within each extracted community. If a node thus becomes isolated, it is also removed. Note that this naturally deals with possibly overlapping communities~\cite{PDFV05} and network hierarchical structure~\cite{CMN08,CMN06}. Hence, same framework might be preferable in different scenarios.

Pool of candidate communities can be generated by directly optimizing~\eqref{W}, or by applying some community detection algorithm~\cite{For10,Sch07}. Due to simplicity, we adopt the latter. Thus, at each step, candidate pool consists of communities identified by the algorithm proposed in~\cite{RB07}. Although not the best community detection approach in the literature~\cite{LF09b,For10}, it is not limited to only community links, which is in our case essential.

\begin{figure}[t]
\includegraphics[width=1.00\columnwidth]{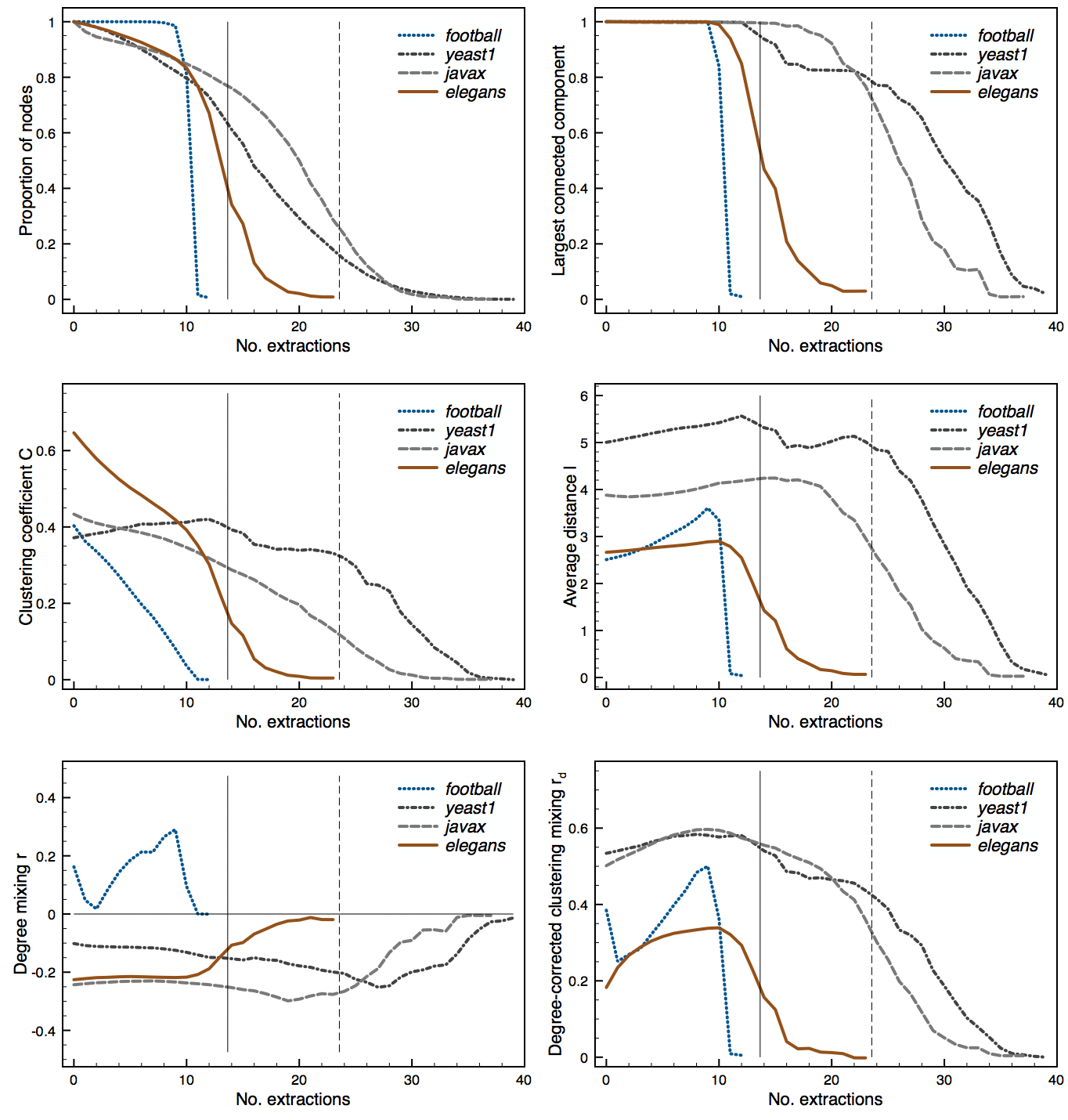}
\caption{\label{fig_ext}(Color online)~Common statistics during community extraction from different real-world networks. (Results are estimates of the mean over $100$ runs, while horizontal lines correspond to termination criteria proposed in~\cite{ZLZ11b}.)}
\end{figure}

\figref{ext}~shows common statistics during extraction of communities from different real-world networks (\tblref{rw}). Note that, in the case of non-social networks---\textit{yeast1}, \textit{javax} and \textit{oregon} networks---the number of nodes decreases only gradually, whereas the largest connected components consist of almost all nodes until the network suddenly dissolves (\figrefs{ext}{top}). Structure below communities is thus more complex than commonly presumed and does not consist of merely, e.g., individual links between communities (see~\figref{over}). On the other hand, \textit{football} social network appears to contain only communities.

\figref{ext}~also indicates the termination criteria proposed in~\cite{ZLZ11b} for \textit{javax} and \textit{elegans} networks. However, one can observe notable changes in network properties even before the latter. As dense modules are being extracted from the network, $r$ and $C$, obviously, decrease (\secref{intro}). Still, $r$ decreases only until pronounced communities are found, and starts to increase, when the rest of the structure is bothered. Similarly, one can observe initial increase in $r_d$, indicating better separated structural modules (see~\secref{mod}), until $r_d$ decreases rather quickly. Due to absence of communities, the algorithm identifies larger groups of isolated nodes, and thus the size of the network also drops suddenly. Any modules beyond communities are therefore characterized by low clustering, degree disassortativity and clustering assortativity.

Observe also that average distance between the nodes denoted~$l$~\cite{AMO93}---the number of links in the shortest path---increases only slightly during extraction of communities. Hence, structure below communities in fact provides efficient global navigation throughout the network, whereas communities improve navigation only locally. Thus, $l$ is only slightly higher in the network after extraction. Similar effect was famously observed by~\citet{Gra73} as the strength of weak ties---links below communities---and the weakness of strong ties---links within communities---and also by other authors more recently~\cite{CRSZZ10,GRMPE12}.

\figref{over}~shows the networks obtained after applying the proposed framework. Communities are extracted as long as the largest connected component consists of more than $99\%$ of the nodes. (Criteria based on, e.g., $r_d$ might be more appropriate for larger networks.) Note that although communities span over large regions of these networks, much of the structure remains unexplained.

We here restate the structural-world conjecture given in~\secref{intro}. Networks can be seen as composed out of two layers. Below, the structure is characterized by functional modules based on the roles played by the nodes withing the underlying systems (see~Secs.~\ref{sec_exp},~\ref{sec_cora}). As shown in~\secref{mod}, this can explain observed clustering, degree disassortativity and efficient global navigation. Above, networks consist of communities that are overlaid over functional modules, and coincide with some assortative property of the nodes~\cite{MSC01,NG03}. Communities increase the overall clustering and degree assortativity in the network, and provide for efficient local navigation.

Distinction between layers is merely artificial, adopted to ease the comprehension. Although this naturally relates the conjecture to the concepts of layered or coupled networks~\cite{MRMPO10,GZXQJHLZ11,GBSH12}. Also, as already stressed in~\cite{SB12u}, semantics behind the links of the two layers necessarily differ. (For incorporation of semantics see, e.g., \cite{LVR10}.) Since assortativity coexists with transitivity~\cite{FFGP11}, a relation that produces communities cannot generate, e.g., disconnected functional modules. Thus, different structural modules are expected only in heterogeneous networks. Nevertheless, most networks are heterogeneous~\cite{SB12u}.

Despite all claims in the paper being investigated thoroughly, we still pose the above as a conjecture. The reason is that low clustering and degree disassortativity characterizing functional modules are already expected properties of, e.g., scale-free networks~\cite{PN03,NP03}. More precisely, if a network is reduced to a simple graph, and the largest degree is at least of order $\sqrt{n}$, the network is likely to be degree disassortative~\cite{MSZ04} (although only a part can be accounted for~\cite{PN03}). Moreover, heterogeneous scale-free networks with maximal entropy are degree disassortative networks~\cite{JTMM10}, whereas disassortativity also results from high branching or low clustering~\cite{Est11}. On the other hand, networks are expected to be highly clustered only in the case of some explicit process that introduces transitivity~\cite{NP03,New03b}. Hence, preferential attachment model~\cite{BA99} that generates scale-free networks can already interpret at least degree mixing and clustering below communities.

However, the above explains network dynamics only at a level of individual nodes, or gives a macroscopic description of the system. Functional modules, together with structural-world conjecture, provide a mesoscopic view. For example, under scale-free model of~\citet{BA99}, nodes preferentially link to hub nodes. The same effect can be encountered under structural-worlds, while functional modules give further knowledge about which hubs are likely to be linked simultaneously. The conjecture thus extends the scale-free phenomena. Similarly, small-world model of~\citet{WS98} demonstrates that introduction of long-range links into else highly clustered network significantly decreases the distances between the nodes. Again, functional modules describe how these links are distributed throughout the network, while high clustering is a result of overlaid communities. Structural-world conjecture thus encloses scale-free and small-world phenomena~\cite{WS98,BA99} by explaining network dynamics through its mesoscopic structures---functional modules and communities.

Similar characteristics of real-world networks have most notably been investigated in the case of Internet~\cite{PVV01,MSZ04}, software systems~\cite{SB12u,SB11s}, web graphs~\cite{FFGP10,SB12u} and biological networks~\cite{MS02,PSR10}, while authors have also considered ensembles of graphs~\cite{JLHJZY07,BW12} and directed networks~\cite{FFGP10}. \citet{HL11}~have already observed an apparent dichotomy in degree mixing of biological networks, which is very similar to our structural-worlds. However, their comprehension of the phenomena was limited to communities. Authors have also analyzed robustness~\cite{Pei11a} and resilience~\cite{Pei12}, and dynamical processes including spreading~\cite{SB12a} and communication flows~\cite{PT11a}. Finally, \citet{PB07}~have shown that two independent parameters are needed to capture network (dis)assortativity, which nicely coincides with our differentiation between two layers of the structural-world.


\begin{figure*}[t]
\includegraphics[width=1.25\columnwidth]{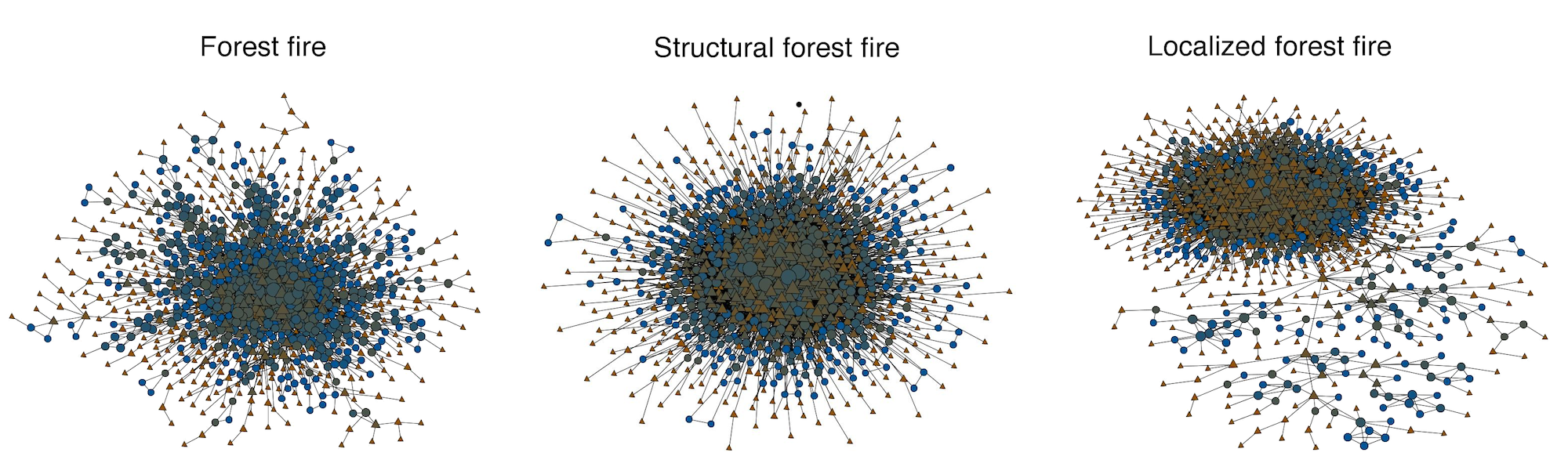}
\caption{\label{fig_ff}(Color online)~Networks containing $1000$ nodes generated with different network models. \ff~model generates degree assortative networks, while those generated with \sff~and \lff~models show degree disassortativity ($r$ equals $0.167$, $-0.189$ and $-0.124$). Clustering $C$ is $0.404$, $0.411$ and $0.338$, while $r_d$ is $0.347$, $0.170$ and $0.285$. (Node symbols are consistent with~\figref{rw}.)}
\end{figure*}

\section{\label{sec_mod}Model}
Proposed network model is based on the burning process of forest fire (\ff) model~\cite{LKF07,LKF05}, which we introduce first. Due to simplicity, the model is presented in the case of undirected networks~\footnotemark[6].

Let $p$ be the burning probability, $p\in [0,1)$ (see below). Initially, network consists of a single node, while for each new node $i$, the burning process proceeds as follows.
\begin{enumerate}[(1)]
\item $i$ chooses an ambassador node $a$ uniformly at random, and links to it.
\item One samples $x_p$ from a geometric distribution with mean $\frac{p}{1-p}$. $i$ selects $x_p$ random neighbors of $a$ that were not yet visited, $j_1,\dots,j_{x_p}$, and links to them.
\item One recursively applies step~(2) to each $j_1,\dots,j_{x_p}$, where the latter are taken as ambassadors of $i$.
\end{enumerate}
(If the number of neighbors in step~(2) is smaller than~$x_p$, $i$~selects as may neighbors as it can.) Since each node can be visited at most once, the burning process surely converges. Thus, to generate a network with $n$ nodes, \ff~model repeats the above procedure $n-1$ times. 

\begin{figure}[t]
\includegraphics[width=1.00\columnwidth]{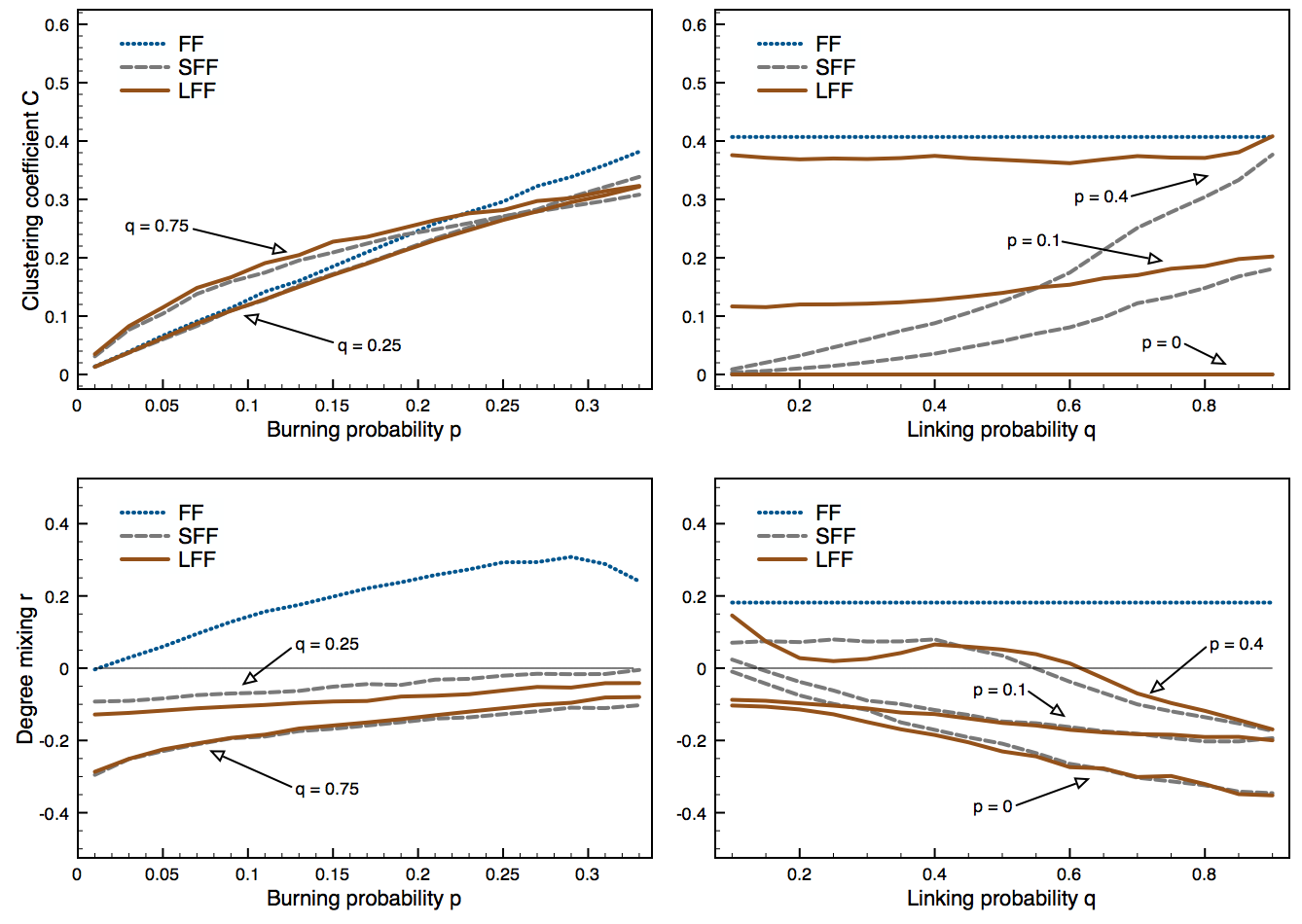}
\caption{\label{fig_pq}(Color online)~Clustering and degree mixing in networks containing $1000$ nodes generated by varying the parameters of different network models. (Results are estimates of the mean over $100$ network realizations.)}
\end{figure}

The model produces densification and shrinking diameter effects observed in temporal real-world networks, while the networks also exhibit skewed degree distribution, short distances between the nodes and community structure~\cite{LKF07,LKF05}. Hence, the model generates networks with high clustering and degree assortativity (see~\figref{pq}).

It also has a natural interpretation in, e.g., citation networks. Burning process mimics the author of a paper including references into bibliography. Author first reads a related paper, or selects a paper that triggered the research, and includes it into bibliography (step~(1)). Author then considers bibliography of the latter, or additional resources, for other related papers (step~(2)). Some of thus discovered papers are further considered and cited, while the author continues similarly as before (step~(3)). Despite the above, \ff~model fails to reproduce the properties observed in citation networks.

Note that described process implicitly assumes that authors read, or at least consider, all the papers they cite. However, this is indeed not the case. For example, seminal work on random graphs conducted by Erd\"{o}s and R\'{e}nyi~\cite{ER59} is perhaps among most widely cited papers in the network science literature. Although, presumably, only a small number of authors have actually read the paper. As the work is widely discussed elsewhere, most authors have just copied the reference from another paper. On the other hand, authors also do not cite all papers they read, although directly related to their work. The latter can be simply due to page limitations or, in the case of papers that appear at about the same time, related paper can limit, question or even contradict the work of the author. Nevertheless, such paper would be read thoroughly, while, presumably, many of its references will be further considered and also cited. 

Examples suggest that the papers that authors read or cite are selected based on two, not necessarily dependent, processes. We thus propose structural forest fire (\sff) model that adopts the same burning procedure as \ff~model to traverse the network, whereas links are formed according to another independent process.

Let $q$ be the linking probability, $q\in [0,1)$ (see below). Initially, network consists of a single link, while for each new node $i$, the model proceeds as follows.
\begin{enumerate}[(1)]
\item $i$ chooses an ambassador node $a$ uniformly at random.
\item One samples $x_p$ from a geometric distribution with mean $\frac{p}{1-p}$. $i$ selects $x_p$ random neighbors of $a$ that were not yet visited, $j_1,\dots,j_{x_p}$.
\item One samples $x_q$ from a geometric distribution with mean $\frac{q}{1-q}$. $i$ selects $x_q$ random neighbors of $a$ that were not yet linked, $l_1,\dots,l_{x_q}$, and links to them. 
\item One recursively applies step (2) to each $j_1,\dots,j_{x_p}$, where the latter are taken as ambassadors of $i$.
\end{enumerate}
(Details are the same as above.~\footnotemark[7]) Again, the process converges, whereas the entire procedure is repeated $n-2$ times. Step~(3) ensures that no multiple links are formed. 

Denote $v$ to be mean number of ambassador nodes visited within the burning process, and let $p<0.5$. Then,
\begin{eqnarray}
v & \leq & \sum_i\left(\frac{p}{1-p}\right)^i \leq \frac{1-p}{1-2p},
\label{eq_v}
\end{eqnarray}
while the expected degree in the network is
\begin{eqnarray}
k & \leq & \frac{2vq}{1-q}.
\label{eq_k}
\end{eqnarray}
Although valid only in the limit of large~$n$, the bounds appear rather tight for small enough $p$ and $q$ (see~\secref{cora}).

Note that, as node does not necessarily link to the ambassador node, it will fail to form any link---become isolated---with probability $(1-q)^v$. Although the latter is negligible in most cases, this is not the case when one models very sparse networks that imply smaller $p$ and $q$ (e.g., power grids). Isolated nodes are else a common property of real-world networks. However, in practice, these are often ignored in the analysis, or the network is even reduced to the largest connected component.

To obtain a connected network with desired number of nodes, one could simply repeat the burning procedure until the property is reached (for other alternatives see~\cite{Vaz03,LKF07}). However, according to the above interpretation, isolated nodes can be seen as authors who failed to find any paper that is indeed similar to their own. The latter can indicate seminal work. In that case, an author would, presumably, attempt to relate the work with existing literature as best as possible. As such process is better imitated by \ff~model, for the analysis here, a node that remains isolated in the final network is re-incorporated according to the dynamics of \ff~model (with same $p$). This ensures a network with $n$ nodes. (Note that the above refers to only $(1-q)^v(n-2)$ nodes.)

\sff~model can generate networks that exhibit most properties observed in real-world networks (see~\figref{ff}). Clustering assortativity $r_d$ introduced in~\secref{mix} is usually around $0.2$, which nicely coincides with, e.g., \textit{elegans} metabolic network that exhibits $r_d=0.183$. However, most other networks in~\tblref{rw} have $r_d$ much larger than that. By observing these networks in~\figref{rw} on can realize that most of them show an interesting phenomena. Communities of highly clustered nodes are localized in certain regions, although these may be scattered across the networks (e.g., \textit{power}, \textit{yeast1} and \textit{javax} networks). (Localization phenomena appears to relate to some property that may vary between the underlying systems.)

We thus also analyze a small variation of the model denoted localized forest fire (\lff) model. Due to simplicity, localization in the model is realized by using isolated nodes. When such node is re-incorporated into the network, the ambassador in step~(1) is chosen only between isolated nodes, and not between all the nodes in the network as originally (omitted for first node). As isolated nodes are re-introduced using \ff~model, communities of highly clustered nodes emerge, while the above variation forces them to localize in certain regions of the network.

\begin{figure}[b]
\includegraphics[width=1.00\columnwidth]{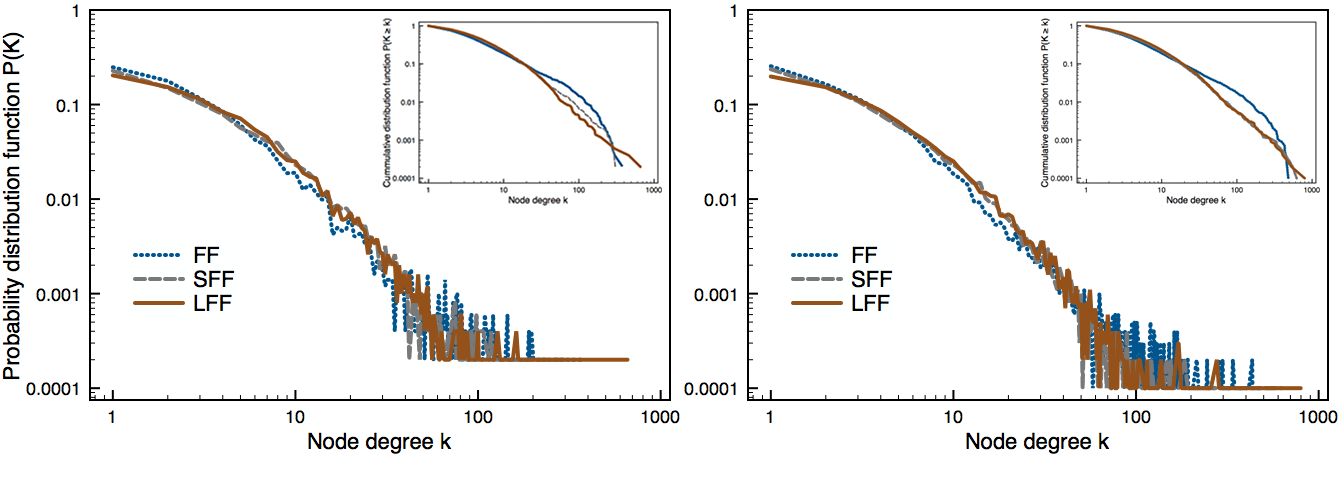}
\caption{\label{fig_k}(Color online)~Degree distributions of networks containing $5000$ and $10000$ nodes generated with different network models (left and right, respectively). Scale-free exponents $\alpha$ for smaller networks generated with \sff~and \lff~models are $2.61$ and $2.85$, while $2.82$ and $2.88$ for larger ones. (Power-law $p_k\propto k^{-\alpha}$ is a plausible fit at $p\mbox{-value}=0.1$ in all cases~\cite{CSN09}.) Average degrees range from $8.1$ to $8.7$, and thus coincide with density of real-world networks~\cite{LJTBH11,BSB12}.} 
\end{figure}

\lff~model indeed generates networks with higher clustering assortativity, where $r_d$ is around $0.3$ in most cases. Localization can also be clearly observed in~\figref{ff}, however, since our intention is merely a statistical analysis of the localization effect, \lff~model is not expected to generate networks that would visually resemble real-world networks (like \ff~and \sff~models). Results further indicate that $r_d$ measures the overlap between the nodes with high and low clustering, or, more precisely, how well are the communities separated from the rest of the network. Note also that all models still seriously underestimate $r_d$.

Both \sff~and \lff~models generate networks that show scale-free degree distributions (\figref{k}) and small-world phenomena (results omitted). In~\figref{pq} we also analyze network clustering and degree mixing by varying the probabilities $p$ and $q$. $p$ governs clustering in the network for all forest fire models, where clustering increases monotonically with $p$. Also, $p$ increases degree assortativity in the case of \ff~model, whereas $p$ has little effect on degree mixing in the case of \sff~and \lff~models. However, degree mixing in the models is governed by~$q$, where degree disassortativity increases with $q$. $q$ also does not affect the clustering for \lff~model.

The above can be summarized as follows. Notice that, due to a special treatment of isolated nodes, \lff~model is in fact identical to \ff~model for $q=0$~\footnotemark[6]. On the other hand, the model generates bipartite networks for $p=0$. (Or $k$-partite networks if the initial network is a full graph on $k$ nodes.) Hence, for large $p$ and small $q$, generated networks experience large clustering, degree assortativity and clear community structure. While, for small $p$ and large $q$, networks show lower clustering and high degree disassortativity. Furthermore, such networks also contain pronounced functional modules, since multi-partite networks are the most prominent examples of the latter. In other words, $p$ and $q$ vary the number of nodes on each layer of structural-world networks (\secref{mix}), while one can thus generate clustering and degree mixing that resembles almost arbitrary network in~\tblref{rw}~\footnotemark[8].

Although not directly discussed above, one can imagine that copying the links of ambassador node---linking to its neighbors---within the proposed models produces functional modules. We stress that the key factor here is that the node does not necessarily link to the ambassador itself. The latter is also the cause for degree disassortativity, whereas linking to the ambassador gives degree assortative networks in any case. Still, degree mixing coefficient $r$ should not be taken as a measure of how pronounced the functional modules are, but is rather related to the difference in the size of the dependent modules. One can expect larger modules in networks that show greater degree disassortativity, whereas low clustering, alongside large average degree, is perhaps the most prominent indicator of functional modules in a network.

Note that functional modules are an implicit result of our models, while this is rather different from some other techniques in the literature. For example, blockmodels~\cite{WBB76}, which were widely studied in social networks literature in the past, can generate networks with arbitrary structural modules. However, one has to specify how different modules are interlinked between, which is often hard in practice~\cite{KN11a,SB12u}.

Other authors have also proposed models very similar to ours~\cite{HK02,LC03,KOSKK07,ZRWZG07}. However, these either do not adopt the burning process to traverse the network~\cite{Vaz03,MAF08} or the model necessarily links the nodes with their ambassadors~\cite{WH09,RCS11}, which results in degree assortativity. Nevertheless, the main difference in these models is that the set of the nodes that are linked with the concerned node is always a subset of the nodes that are visited. For the models proposed here, these two sets can intersect arbitrarily, while they can also be disjoint. The latter in fact produces rich dynamics observed above.

Last, we also note that, although the interpretation of the models was limited to citation networks, the same process accurately resembles the dynamics in many other networks. For example, in the case of software networks, the process exactly mimics the actions of a developer that is adding classes of code into unfamiliar project. Similarly, in air transportation networks, an airport might offer the same flights as a nearby airport---and would thus copy its links---while it will, presumably, not offer a flight towards the latter. Finally, the same process can also be undertaken by a person that is being familiarized with friends of an acquaintance. Although the person does not become a friend of the acquaintance, he or she may still befriend with some of the acquaintance's friends.


\begin{figure*}
\begin{eqnarray}
g_i & = & \argmax_g \left[\left.\nu_g\sum_{j\in\neigh_i} f_j\cdot\delta(g_j,g)\right. + \left.(1-\nu_g)\sum_{j\in\neigh_i}\sum_{l\in\neigh_j\setminus\neigh_i} \tilde{f}_l/k_j\cdot\delta(g_l,g)\right.\right]
\label{eq_gp}
\end{eqnarray}
\end{figure*} 

\section{\label{sec_alg}Algorithm}
Despite the discussion in~\secref{mix}, proposed algorithm directly partitions the network into structural modules without first extracting communities. The algorithm is based on a label propagation framework proposed in~\cite{SB12u}, which we briefly introduce below. Due to simplicity, the framework is presented in the case of simple graphs.

Denote $\neigh_i$ to be the set of neighbors of node $i$ and let $g_i$ be an unknown module (or group) label. Furthermore, let $\g_i$ be the set of nodes sharing label $g_i$. Propagating labels between the nodes was first introduced for community detection. \citet{RAK07}~have proposed a simple algorithm that exploits the following procedure. Initially, each node is labeled with a unique label, $g_i=i$. Then, at each iteration, a node adopts the label shared by most of its neighbors (updates occur in a random order). Hence,
\begin{eqnarray}
g_i & = & \argmax_g \sum_{j\in\neigh_i} \delta(g_j,g),
\label{eq_lp}
\end{eqnarray}
where $\delta$ is the Kronecker delta (ties are broken uniformly at random~\footnotemark[9]). Due to many links within communities, relative to the number of links towards the rest of the network, nodes in communities form a consensus on some label after a few iterations. Thus, when an equilibrium is reached, $\{\g\}$ contains communities that are clearly depicted in the network structure. Note that~\eqref{lp} is actually equivalent to a simple kinetic Potts model~\cite{TK08}.

Due to extremely fast structural inference of label propagation, the algorithm exhibits near linear complexity~$\mathcal{O}(m^{1.2})$, while the expected number of iterations on a network with a billion links is only $113$~\cite{SB11d}.

The basic algorithm can be further improved by also applying node preferences~\cite{LHLC09}. Preferences adjust the strength of propagation from certain nodes (e.g., hubs), and thus force the propagation process towards more desirable partitions. Let $f_i$ be a preference of node $i$. Then,
\begin{eqnarray}
g_i & = & \argmax_g \sum_{j\in\neigh_i} f_j\cdot\delta(g_j,g).
\label{eq_dp}
\end{eqnarray}

Note that preferences can be set to an arbitrary node property (e.g., degree). We adopt the same preferences $f$ and $\tilde{f}$ (see~\eqref{gp}) as in~\cite{SB12u}, whereas detailed description is omitted here (due to space limitations). However, $f$, $\tilde{f}$ are in fact composed out of two factors. First factor corresponds to balanced propagation~\cite{SB11b} that, at each iteration, decreases the propagation strength from the nodes that are considered first, and increases the strength from the nodes that are considered last. This counteracts for the randomness introduced through random update orders, which stabilizes the propagation process~\footnotemark[10].

Second factor corresponds to defensive propagation~\cite{SB11d} that further increases the propagation strength from the core of each $\g$ or, equivalently, decreases the strength from its border. The latter forces the algorithm to gradually reveal the network structure and improves its detection strength in real-world networks~\cite{SB11d,SB10k}.

The algorithm in~\eqref{dp} is comparable to current state-of-the-art in community detection, while \citet{SB12u} have extended the same principle to also functional modules. Rather than propagating the labels between the neighboring nodes, one propagates the labels between the nodes at distance two---through common neighbors. Since nodes in functional modules share many common neighbors, similarly as before, they form a consensus on some particular label. Thus, when the process unfolds, $\{\g\}$ contains most pronounced functional modules in the network. Generalization is identical to a standard propagation on a network with the same set of nodes, while the links represent (link-disjoint) paths of length two between the nodes in the original network.

The algorithm for functional modules is shown in the right-hand side of~\eqref{gp}. $\tilde{f}_i$ is preference of node $i$ (see above), whereas $k_j$ is a normalization. Since the sum in~\eqref{dp} has $k_i$ terms, whereas the sums in the right-hand side of~\eqref{gp} contain up to $\sum_{j\in\neigh_i}k_j$ terms, this introduces biases in the detection of modules. Dividing each term by $k_j$ aggregates the contributions at each neighbor of $i$, which makes all sums proportional to $k_i$.

\eqref{gp}~shows complete algorithm for detection of arbitrary structural modules denoted generalized propagation~\cite{SB12u}. Left-hand side is the same as~\eqref{dp}, while $\nu_g$ are module dependent factors that represent adopted network modeling, $\nu_g\in[0,1]$. Setting $\nu_g=1$ for all $g$ is identical to community detection algorithm in~\eqref{dp}, whereas, for all $\nu_g$ equal to zero, the algorithm reveals only functional modules. When $\nu_g=0.5$, identified modules depend on community and functional links.

General propagation framework in~\eqref{gp} can detect communities and functional modules even when only weakly depicted in the structure of the network~\cite{SB12u,SB11g}. However, module factors~$\nu_g$ have to be defined accordingly (see~\appref{mod}). Note that $\nu_g$ are set apriori, when one initializes node labels $\{g\}$. As most labels disappear during propagation, one only has to provide sufficient number of labels with $\nu_g\approx 1$ in regions of the network, where communities could exist, and enough labels with $\nu_g\approx 0$ in regions, where one expects functional modules.

Network modeling proposed in~\cite{SB12u} was based on conductance~\cite{Bol98}, while the algorithm in~\cite{SB11g} uses clustering coefficients $c$ and $C$~\cite{WS98}. Due to the analysis in~\secref{mix}, we propose a much simpler model based on degree-corrected clustering coefficients $d$ and $D$~\cite{SV05}.
\begin{subnumcases}{\label{eq_mod}\nu_{g}=}
\label{eq_mod1}
1 & for $D\geq\tau \wedge d\geq\tau$\\
\label{eq_mod0}
0 & for $D<\tau \wedge d<\tau$\\
\label{eq_mod5}
0.5 & otherwise
\end{subnumcases}
Parameter $\tau$ is expected clustering in a corresponding random graph, where Erd\"{o}s-R\'{e}nyi~\cite{ER59} graph with clustering $p_r$ and configuration model~\cite{MR95,NSW01} with $p_c$ are the most obvious choices (\secref{mix}). For the analysis here, we set $\tau$ to $p_c$, to reveal modules that go beyond (scale-free) degree distributions (if not stated otherwise).

\eqref{mod}~should be seen as follows. Since most real-world networks have $D\geq\tau$, the algorithm identifies communities in regions with high clustering (\eqref{mod1}). \eqref{mod0}~properly models, e.g., bipartite networks, where functional modules are revealed in regions with very low clustering. Otherwise, $\nu_g=0.5$~(\eqref{mod5}).

Note that, according to structural-world conjecture, functional modules also emerge below communities, in regions with high clustering. The model in~\eqref{mod} thus apparently ignores most functional modules in the network. However, different modules commonly become obscure in the presence of communities. Let a network contain a complete $k$-partite subgraph on $n$ nodes. Obviously, the network contains clear functional modules. Still, when $k$ goes to $n$, the corresponding subgraph actually becomes a clique, and thus a well defined community. Similar effect occurs, when the size of subgraph decreases or, equivalently, $n$ goes to $k$. Hence, functional modules often cannot be revealed alongside communities.

The above is directly related to the following question. Assume that networks truly contain functional modules as structural-world predicts; how that there exist numerous studies in the literature, where authors have identified communities in almost any type of real-world networks? Analysis in~\secref{exp} shows that community detection algorithms commonly identify dependent functional modules as a single group of nodes. Nonetheless, the latter can still be a well defined community. Actually, some of the communities in the famous \textit{football}~social network (\tblref{rw}) are in fact multi-partite graphs (see~\figref{afl}).

Although this indicates a rather undesirable behavior of community detection algorithms, it can be employed for detection of other modules. The model in~\eqref{mod} thus first tries to identify dependent functional modules as communities, which are then (possibly) refined into functional modules. Let $\{\g\}$ be the initial partition revealed by the algorithm. For each $\g$, the algorithm is further applied to a subnetwork induced by the nodes in~$\g$. As this refinements proceed recursively, an entire sub-hierarchy of modules is obtained. Similarly, one can reveal a super-hierarchy by applying the algorithm to a super-network induced by $\{\g\}$ (agglomeration). Here (super-)nodes represent modules that are linked, when a link between their nodes also exists in the network. Final result of such divisions and agglomerations is a complete hierarchy of modules denoted $\h$~\cite{CMN06, CMN08} (see~\figref{afl}).

Leafs of $\h$ represent nodes in the network, while inner nodes $\{\I\}$ correspond to modules $\{\g\}$ that were obtained over several applications of the algorithm. Let $\I$ correspond to module $\g$ and let $\I_1,\dots,\I_t$ be the ancestors of $\I$. Furthermore, let $\h_{\I}$ be a sub-hierarchy rooted at $\I$. Thus, leafs of $\h_{\I}$ are nodes in module $\g$, while $\g_1,\dots,\g_t$ is a partition of the subnetwork induced by $\g$. Let each inner node $\I$ (or module $\g$) also be associated with value $\theta$, $\theta\in[0,1]$, which is defined as the probability that two nodes in $\g_i$ and $\g_j$ are linked, $i\neq j$. Let there be $m$ such links in $\g$ and denote $s$ to be the number of all possible links, $s=\sum_{i<j}|\g_i| |\g_j|$. Then, $\theta=m/s$. (If ancestors of $\I$ are leafs of $\h$, $\theta$ is just the density of module $\g$.)

Probabilities $\theta$ are in fact maximum likelihood estimators for hierarchy $\h$~\cite{CB90}, while the posterior probability of $\h$---likelihood $\like$ given the network observed---is
\begin{eqnarray}
\like(\h) & = & \prod_{\I\in\h} \left(\theta_{\I}\right)^{m_{\I}} \left(1-\theta_{\I}\right)^{s_{\I}-m_{\I}}.
\label{eq_L}
\end{eqnarray}
Analysis in~\secref{exp} reports log-likelihoods \mlogL, where smaller values are better. Note that $\log\mathcal{L}$ of $\{\g\}$ is the entropy of the corresponding blockmodel~\cite{Pei11b}.

When a network contains only communities, initial partition $\{\g\}$ would commonly already identify them. Thus, $\{\g\}$ should not be further refined, which requires an extra criteria. A promising approach is to refine only very sparse modules $\g$ that clearly do not coincide with the definition of a community, e.g., when a subnetwork induced by $\g$ have $D<\tau$ (results omitted). For the analysis here, $\g$ is refined when $|\g|>3$, while only refinements with $\like(\h_{\I})>\like(\g)$ are accepted---$\h_{\I}$ is the revealed hierarchy for $\g$, and $\like(\g)$ is the likelihood of a hierarchy with a single (inner) node and leafs from $\g$.

Proposed algorithm thus constructs an entire hierarchy of modules $\h$ and is denoted hierarchical propagation (\hpa) algorithm. When only a partition of the network is required, modules represented by the bottom-most (inner) nodes in $\h$ are reported (no agglomerations are needed). We also consider an algorithm without module refinements, which is, for consistency with~\cite{SB12u}, denoted generalized propagation (\gpa) algorithm.

Complexity of \hpa~algorithm is near ideal. Each propagation of labels (\eqref{gp}) requires $\mathcal{O}(km)$, where $m$ is the number of links and $k$ the average degree. The number of iterations required for the propagation process to converge can be, according to~\cite{SB11d} and above analogy, estimated to $\mathcal{O}((km)^{0.2})$. Since the number of module refinements is usually negligible, the total complexity becomes $\mathcal{O}((km)^{1.2})$. When an entire hierarchy is required, the latter refers to a single level. (For the analysis here, we limit the maximum number of iterations to $100$.)

\hpa~algorithm is first validated on synthetic benchmark networks, and applied to different real-world networks (\secref{exp}). Next, the algorithm is adopted for the analysis of a citation network (\secref{cora}), where communities have been extracted apriori. Comparison with $12$ state-of-the-art algorithms is conducted in~\appref{soa}, while \appref{mod}~also analyses different network models (\eqref{mod}).


\begin{figure*}[t]
\includegraphics[width=1.875\columnwidth]{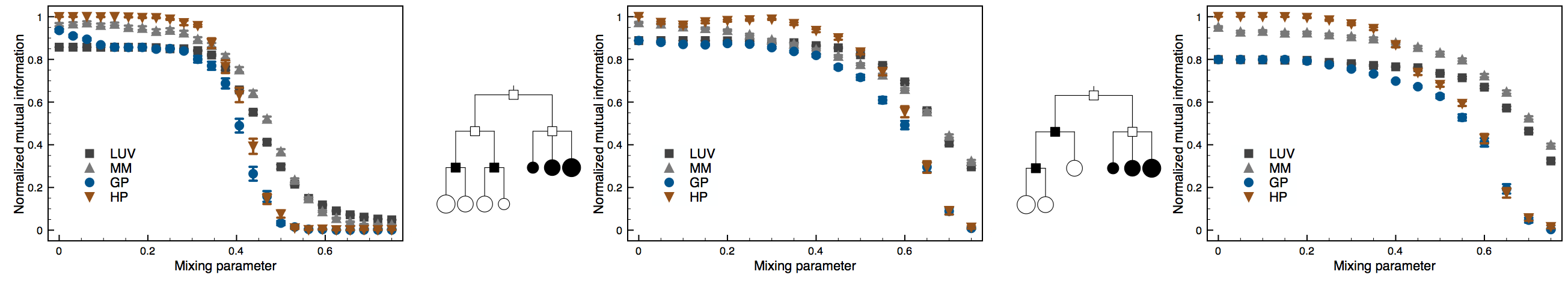}
\caption{\label{fig_syn}(Color online)~Comparison of module detection algorithms on~\GNT, \HNV~and \HNX~synthetic networks (left to right, respectively). (Results are estimates of the mean over $100$ network realizations, while bars show standard error of the mean.)}
\end{figure*}

\section{\label{sec_exp}Structural modules}
Results in the following (and in Apps.~\ref{app_soa},~\ref{app_mod}) are reported in terms of \logL~(\eqref{L}), normalized mutual information~\cite{DDDA05} (\nmi) and adjusted rand index~\cite{Mei07} (\ari). \nmi~equals to the mutual information of the true network partition, and partition of modules revealed by the algorithm, that is normalized by their entropies, $\mathrm{NMI}\in[0,1]$. \ari~measures the fraction of node pairs that are classified the same in both partitions, subjected to the expected value in a null model, $\mathrm{ARI}\in[0,1]$. For both measures, identical partitions experience $1$, while the expected value for independent partitions is zero.

Proposed \hpa~algorithm is compared against a variation without module refinements denoted \gpa~algorithm (\secref{alg}), and a variation limited to merely communities denoted \cpa~algorithm (see~\appref{mod}). We also consider two other approaches. First is a community detection algorithm known as Louvain method~\cite{BGLL08} that optimizes modularity~\cite{NG04} by multi-level aggregation~(\luv\space algorithm). Second is a (general) structural module detection algorithm, which fits a predefined mixture model using expectation-maximization~\cite{DLR77} technique~\cite{NL07} (\mm~algorithm). Among $13$ state-of-the-art algorithms in~\appref{soa}, both represent the second best approach in their category (after that in~\cite{RB08} and \hpa~algorithm).

\mm~algorithm demands the number of modules to be known beforehand, which is set to the true value in all cases. (Due to different stopping criteria, results for \mm~algorithm are slightly better than in~\cite{SB12u,SB11g}.)


\subsection{\label{sec_exp_syn}Synthetic networks}
Algorithms are first applied to three synthetic benchmark networks with planted partitions of structural modules. All partitions consist of communities and functional modules, while the structure is controlled by a mixing parameter $\mu$, $\mu\in[0,1]$. When $\mu$ equals zero, all links in the network are placed according to the predefined partition, whereas the structure degenerates with increasing $\mu$.

\figrefs{syn}{left}~plots the results for \GNT~synthetic networks~\cite{PSR10} that are a generalization of a classical community detection benchmark~\cite{GN02}. Networks consist of four modules with $32$ nodes, where two modules are classical communities, while the other two form a bipartite structure of functional modules. Average degree is fixed to~$16$.

Only \hpa~algorithm can accurately reveal the planted structure in these networks. Observe that performance of a community detection algorithm (e.g., \luv~algorithm) is only slightly worse compared to a general module detection algorithm (e.g., \mm~algorithm). The former identifies functional modules in the network as a single group of nodes, however, this is only weakly expressed by standard measures used in the literature~\cite{DDDA05,For10}. As already discussed in~\secref{alg}, \hpa~algorithm first reveals communities in these networks, which are then refined into functional modules (see \gpa~algorithm).

Next, algorithms are applied to two synthetic networks that are based on a hierarchical model~\cite{CMN06, CMN08} (\secref{alg}). Respective hierarchies are shown in~\figref{syn}, where shades of nodes correspond to values of $\theta$ that are set to either $\mu$ or $1-\mu$. For $\mu=0$, $\theta\in\{0,1\}$. Leafs of hierarchies represent actual planted modules with $8$, $16$ or $24$ nodes. Networks thus contain $7$ and $6$ modules, and are denoted \HNV~and \HNX~networks. Note that both networks consist of three communities, whereas functional modules form two bipartite structures in the case of \HNV~networks, and a tripartite structure in the case of \HNX~networks.

Results for \HNV~networks are shown in~\figrefs{syn}{middle}, while \figrefs{syn}{right} plots results for \HNX~networks. Again, only \hpa~algorithm can accurately detect the true structure planted into these networks, while the performance of other algorithms is else similar as above.

\hpa~and \gpa~algorithms are further applied to Erd\"{o}s-R\'{e}nyi~\cite{ER59} random graphs and Barab\'{a}si-Albert~\cite{BA99} scale-free networks that, presumably, contain no structural modules. Number of nodes is fixed to $128$ in both cases, while average degree is varied from $2$ to $32$. When degree exceeds a certain threshold, none of the algorithms reveal any structure in these networks. The transition occurs between $8$ and $16$. The algorithms also do not suffer from the resolution limit problem~\cite{FB07} (results omitted).

We conclude that \hpa~algorithm can detect arbitrary structural modules---communities or functional modules---when they are present in the network. Also, comparative analysis in~\appref{soa} shows that the algorithm outperforms all state-of-the-art approaches considered. As the number of modules is an implicit result of the propagation process employed, the value does not have to be set apriori as for most other algorithms~\cite{NL07,KN11a}. Hence, proposed algorithm represents a promising approach for exploratory analysis of real-world networks.

Next, we adopt \hpa~algorithm for the analysis of real-world networks, with special emphasis on the context of different modules within the underlying systems.


\subsection{\label{sec_exp_rw}Real-world networks}
We first compare the algorithms on four real-world networks from~\tblref{rw}. We adopt three classical social networks---\textit{football}, \textit{karate} and \textit{women} networks---with known sociological classification of the nodes that results from earlier studies~\cite{GN02,Zac77,DGG41}. For \textit{football} and \textit{karate} networks, corresponding network modules are communities that coincide with some assortative property of the nodes, whereas, in the case of \textit{women} network, the partition corresponds to functional modules that represent different roles nodes play within the underlying domain---women and events attended by the former. We also adopt \textit{jung} software network, where nodes represent classes of code that constitute JUNG library~\cite{OFWSB05}. Here software packages decided by the developers are taken as the true network partition. Respective structural modules are thus communities of classes implementing common functionality, and functional modules representing the roles of classes within JUNG project~\cite{SB12u,SB11s} (see below).

The results are shown in~\tblref{com}. Proposed \hpa~algorithm most accurately reveals the true partition in all cases except for \textit{karate} network. More precisely, the algorithm identifies three communities (on average), while sociological partitioning contains two. Nevertheless, partition with three communities is somewhat more consistent with the structure of the network, and thus commonly reported by the algorithms in the literature~\cite{HCZLDF08,MD09}.

\begin{table}[h]
\begin{ruledtabular}
\begin{tabular}{ccccccccc}
\multirow{2}{*}{Network} & \multicolumn{4}{c}{\nmi}  & \multicolumn{4}{c}{\ari} \\
 & \luv & \mm & \cpa & \hpa & \luv & \mm & \cpa & \hpa \\\hline

\textit{football} & $0.876$ & $0.823$ & $0.905$ & $\mathbf{0.909}$ & $0.771$ & $0.683$ & $0.841$ & $\mathbf{0.850}$ \\

\textit{karate} & $0.629$ & $\mathbf{0.912}$ & $0.834$ & $0.866$ & $0.510$ & $\mathbf{0.912}$ & $0.823$ & $0.861$ \\

\textit{jung} & $0.605$ & $0.662$ & $0.650$ & $\mathbf{0.684}$ & $0.269$ & $0.276$ & $0.218$ & $\mathbf{0.280}$ \\

\textit{women} & $0.309$ & $0.825$ & $0.217$ & $\mathbf{0.932}$ & $0.174$ & $0.716$ & $0.119$ & $\mathbf{0.936}$ \\

\end{tabular}
\end{ruledtabular}
\caption{\label{tbl_com}Comparison of module detection algorithms on different real-world networks with known partitioning. (Results are estimates of the mean over $100$ runs.)}
\end{table}

Although structural modules in, e.g., \textit{football} and \textit{women} networks are fundamentally different in many basic module properties, \hpa~algorithm accurately detects the structure that is present in each network. We thus also apply the algorithm to a larger number of real-world networks, to analyze, whether general modules---communities and functional modules---better model the network structure than communities alone. For a fair analysis, we compare modules revealed under the same framework, exemplified by \hpa~and \cpa~algorithms. (For comparison, we also report results for other~approaches.)

\tblref{logLm}~shows \logL~of hierarchies of modules identified with different algorithms (as described in \secref{alg}). We consider $16$ real-world networks from~\tblref{rw} that are reduced to the largest connected components. Observe that general modules better predict the structure present in most types of real-world networks including information (e.g., \textit{javadoc} network), biological (e.g., \textit{yeast2} and \textit{elegans} networks), technological (e.g., \textit{power} and \textit{jung} networks) and, surprisingly, also classical social networks (e.g., \textit{football} and \textit{karate} networks). On the other hand, collaboration networks appear to be the most prominent examples of networks with only community structure  (e.g., \textit{netsci} and \textit{comsci} networks). 

\begin{table}[t]
\begin{ruledtabular}
\begin{tabular}{ccccccc}
\multirow{2}{*}{Network} & \multicolumn{6}{c}{\mlogL} \\
 & \multicolumn{2}{c}{Groups} & \luv & \mm & \cpa & \hpa \\\hline

\textit{netsci} & & & $2622.3$ & & $\mathbf{2024.7}$ & $2152.0$ \\

\textit{comsci} & & & $2399.7$ & & $\mathbf{2011.0}$ & $3326.5$ \\

\textit{football} & $12$ & $1184.3$ & $1138.5$ & $1351.7$ & $1096.0$ & $\mathbf{1095.2}$ \\ 

\textit{karate} & $2$ & $196.3$ & $\mathit{178.0}$ & $195.8$ & $188.2$ & $\mathbf{186.9}$ \\ 

\textit{emails} & & & $25903.6$ & & $\mathbf{24287.7}$ & $24402.8$ \\

\textit{euro} & & & $5954.1$ & & $4202.1$ & $\mathbf{4072.3}$ \\ 

\textit{power} & & & $33874.6$ & & $21132.9$ & $\mathbf{20678.8}$ \\

\textit{javadoc} & $22$ & $23011.5$ & $19943.4$ & $25361.7$ & $19405.7$ & $\mathbf{19346.0}$ \\

\textit{yeast1} & & & $32621.3$ & & $\mathbf{28280.9}$ & $28689.4$ \\

\textit{yeast2} & & & $15766.8$ & & $12826.9$ & $\mathbf{12181.3}$ \\ 

\textit{javax} & $21$ & $16140.7$ & $13728.9$ & $17299.2$ & $13497.5$ & $\mathbf{13063.5}$ \\ 

\textit{jung} & $38$ & $3081.6$ & $2583.8$ & $2997.5$ & $2497.4$ & $\mathbf{2489.6}$ \\ 

\textit{guava} & $4$ & $1418.1$ & $1020.4$ & $1295.8$ & $1012.1$ & $\mathbf{991.3}$ \\

\textit{elegans} & & & $\mathit{8673.5}$ & & $8955.7$ & $\mathbf{8856.9}$ \\ 

\textit{oregon} & & & $\mathit{9351.5}$ & & $\mathbf{9865.6}$ & $10456.6$ \\

\textit{women} & $3$ & $193.3$ & $204.2$ & $\mathit{163.6}$ & $207.4$ & $\mathbf{183.8}$ \\ 

\end{tabular}
\end{ruledtabular}
\caption{\label{tbl_logLm}Analysis of hierarchical structures revealed with module detection algorithms in different real-world networks. (Results are estimates of the mean over at least $100$ runs, while 'groups' corresponds to a known partitioning.)}
\end{table}

\begin{table}[b]
\begin{ruledtabular}
\begin{tabular}{cccccccccc}
\multirow{2}{*}{Network} & \multicolumn{6}{c}{\mlogL~and no. levels} \\
 & Runs & \multicolumn{2}{c}{\cpa} & \multicolumn{4}{c}{\hpa---$p_r$ and $p_c$} & \multicolumn{2}{c}{\cite{CMN06}} \\\hline

\textit{football} & $10^4$ & $1010.9$ & $3$ & $\mathbf{954.8}$ & $\mathbf{5}$ & $1004.1$ & $3$ & $\mathit{884.2}$ & $\mathit{11}$ \\

\textit{karate} & $10^5$ & $174.1$ & $3$ & $\mathbf{172.3}$ & $\mathbf{3}$ & $173.9$ & $2$ & $\mathit{73.3}$ & $\mathit{10}$ \\

\textit{euro} & $10^3$ & $4108.9$ & $6$ & $\mathbf{3883.2}$ & $\mathbf{8}$ & $3924.4$ & $5$ \\

\textit{yeast2} & $10^2$ & $12495.0$ & $6$ & $11611.2$ & $7$ & $\mathbf{11596.4}$ & $\mathbf{4}$ \\

\textit{javax} & $10^2$ & $13020.7$ & $4$ & $12894.1$ & $4$ & $\mathbf{11512.2}$ & $\mathbf{3}$ \\

\textit{jung} & $10^3$ & $2354.5$ & $5$ & $2312.5$ & $4$ & $\mathbf{2272.9}$ & $\mathbf{4}$ \\

\textit{elegans} & $10^2$ & $8734.1$ & $5$ & $8640.9$ & $6$ & $\mathbf{8243.3}$ & $\mathbf{5}$ \\

\textit{women} & $10^4$ & $193.9$ & $2$ & $\mathbf{163.6}$ & $\mathbf{1}$ & $\mathbf{163.6}$ & $\mathbf{1}$ \\

\end{tabular}
\end{ruledtabular}
\caption{\label{tbl_logLp}Analysis of hierarchical structures revealed with module detection algorithms in different real-world networks. (Results are peak values estimated over at least $100$ runs, while last column corresponds to (binary) hierarchies in~\cite{CMN06}.)}
\end{table}

\begin{figure*}[t]
\includegraphics[width=1.50\columnwidth]{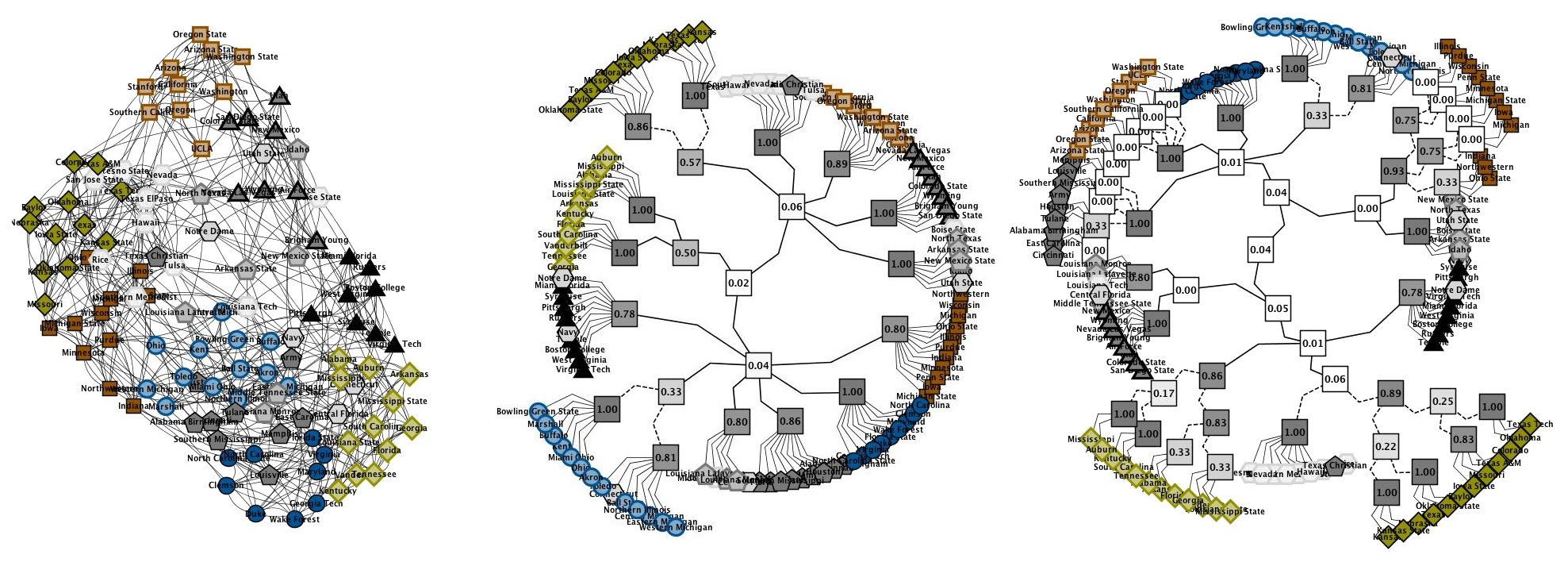}
\caption{\label{fig_afl}(Color online)~Hierarchies of \textit{football} social network revealed with \cpa~and \hpa~algorithms (left-hand and right-hand side, respectively). (Node symbols represent NCAA college (American) football conferences and are consistent among figures. Values in inner nodes of hierarchies equal $\theta$, whereas solid and dashed links correspond to agglomeration and division steps.)}
\end{figure*}

Note that, e.g., \textit{karate} and \textit{elegans} networks are in fact best modeled by considering merely network modularity (see \luv~algorithm). However, the latter is actually due to increased complexity, when one considers hierarchies of general structural modules. For example, assume that the algorithm reveals several functional modules at some level of the hierarchy. In order to adequately model the structure present, dependent functional modules must necessarily be detected as a single group on the next level.

\tblref{logLp}~thus also shows \logL~for the best hierarchies revealed in $8$ real-world networks from~\tblref{logLm}. Despite the discussion above, \hpa~algorithm still most accurately predicts the hierarchical structures in these networks. Results also reveal that $p_c$---expected clustering in a configuration model~\cite{MR95,NSW01}---more adequately distinguishes between different types of modules than $p_r$---clustering of an Erd\"{o}s-R\'{e}nyi random graph~\cite{ER59} (see~\eqref{mod}). Nevertheless, since $p_c$ is merely an upper bound for $C$~\cite{EMB02}, $p_r$ proves to be more appropriate for smaller or very sparse networks (e.g., \textit{karate} and \textit{euro} networks).

\tblref{logLp}~further shows the number of (non-trivial) levels that constitute each hierarchy---height of the hierarchy. Observe that hierarchies of communities generally consist of much larger number of levels than those that include also other structural modules. Since the complexity of the hierarchy increases exponentially with its height, the difference is substantial. \tblref{logLp}~also reports binary hierarchies from~\cite{CMN06} that were reveled by Monte Carlo sampling scheme. Although values of \logL~are better than those obtained with, e.g., \hpa~algorithm, heights of respective hierarchies are again considerably larger.

We conclude that general structural modules more adequately model the mesoscopic structure of networks than communities alone. Next, we consider modules identified in different types of real-world networks in greater detail.

\figref{afl}~shows hierarchies of \textit{football} social network revealed with \cpa~and \hpa~algorithms. Nodes in the network represent college (American) football teams that are linked, when a game was played during the NCAA $2000$ season. Known partitioning of the network corresponds to a division into conferences. Observe that both hierarchies identify conferences at some intermediate level, whereas many are further partitioned into different structural modules. More precisely, although some conferences correspond to a clique in the network, others are in fact multi-partite structures (see~\figrefs{afl}{right}). The latter can be directly related to the schedule of the games played, and thus the role of different teams during the $2000$ season. Hence, despite the fact that the network represents a classical benchmark for community detection, hierarchy of merely communities fails to identify most of the structure present (see~\figrefs{afl}{left}).

\figrefs{jung}{right}~also shows hierarchy of \textit{jung} software network (see above) revealed with \hpa~algorithm. Modules again coincide with the known classification of the nodes---packages of software classes---whereas different modules show characteristic features of JUNG project. For example, classes implementing the same functionality (e.g., graph implementations) commonly correspond to a community in the network, while classes with the same role within the project (e.g., parsers or plug-ins), which depend on a common set of other classes, are usually expressed as functional modules. Again, much of the functionality of JUNG library would remain obscure under the framework limited to communities (see also~\cite{SB12u}).

\begin{figure}[b]
\includegraphics[width=1.00\columnwidth]{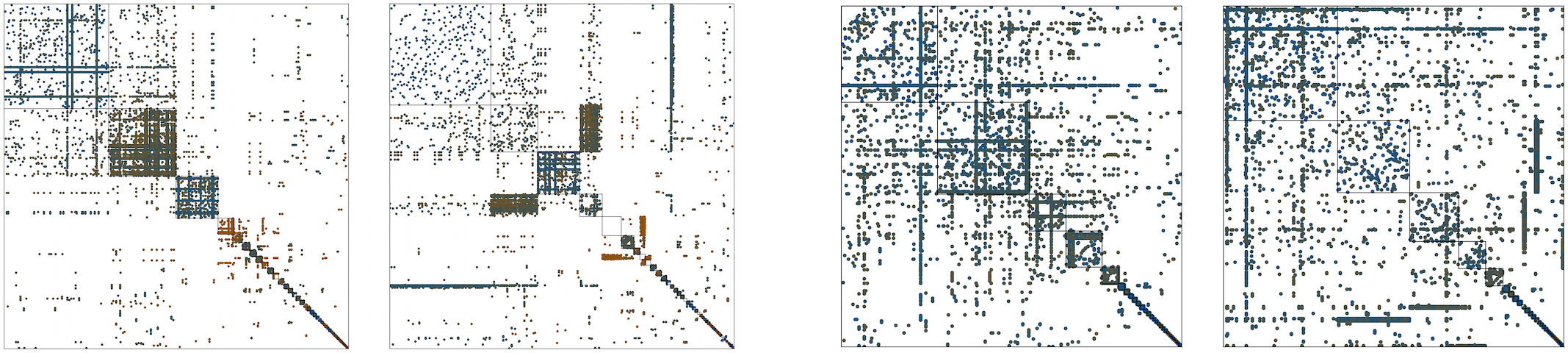}
\caption{\label{fig_block}(Color online)~Structural modules---blockmodels---of \textit{javax} software and \textit{elegans} metabolic networks revealed with \cpa~and \hpa~algorithms (left-hand and right-hand side, respectively). (Dots represent links, whereas shades correspond to average $d$ at link ends and range between $0$ ({\color{mocha}mocha}) and $1$ ({\color{blue}blue}). Dots are enlarged five times for better visibility.)}
\end{figure}

\begin{figure*}[t]
\includegraphics[width=1.66\columnwidth]{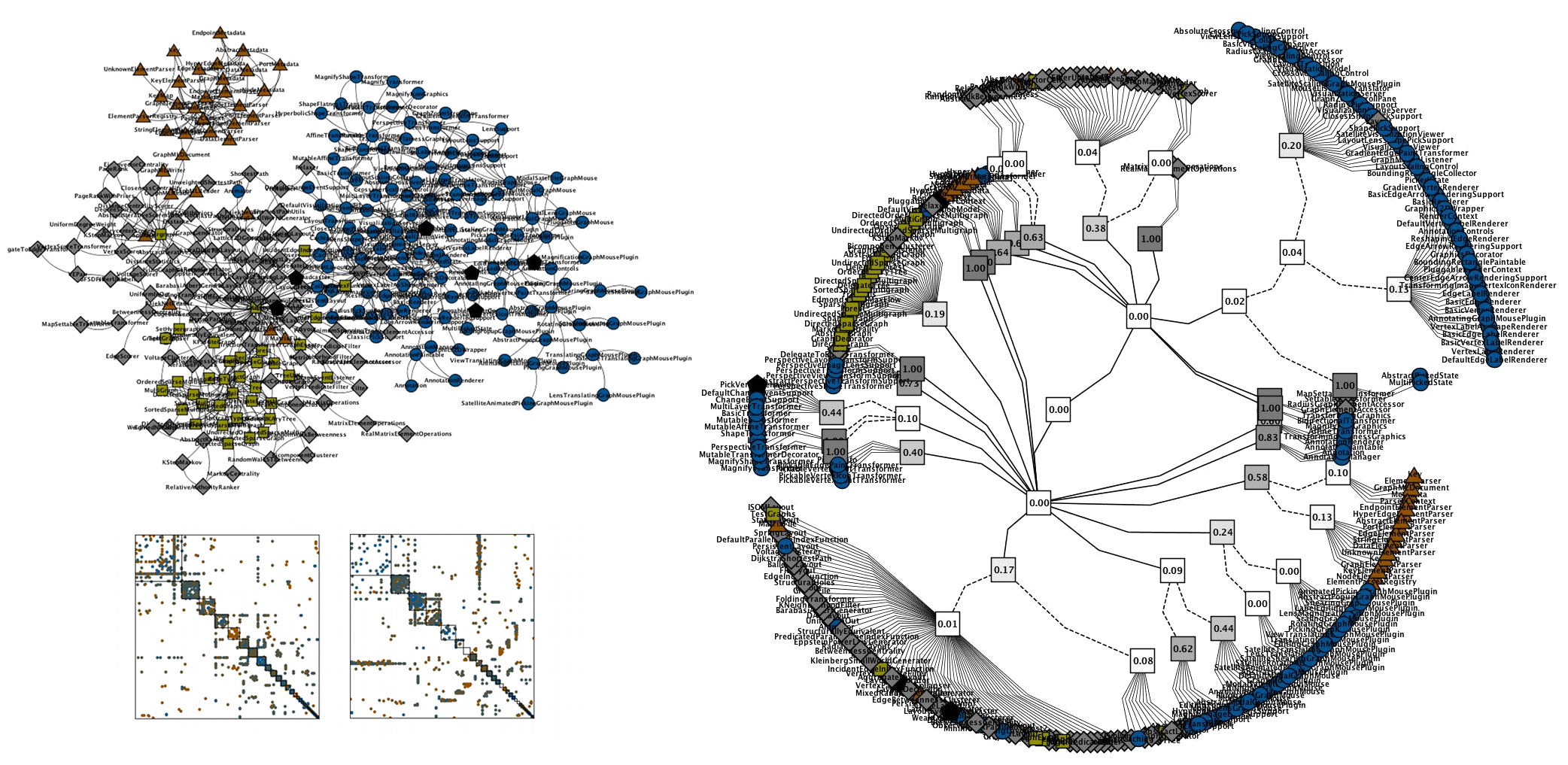}
\caption{\label{fig_jung}(Color online)~(left)~Structural modules---blockmodels---of \textit{jung} software network revealed with \cpa~and \hpa~algorithms (left-hand and right-hand side, respectively). (Symbols are consistent with~\figref{block}.) (right)~Hierarchy of \textit{jung} software network revealed with \hpa~algorithm. (Node symbols represent high-level packages of JUNG library~\cite{OFWSB05}---\texttt{jung.visualization} ({\color{blue}circles}), \texttt{jung.io} ({\color{mocha}triangles}), \texttt{jung.graph} ({\color{asparagus}squares}) and \texttt{jung.algorithms} ({\color{gray}diamonds})---while hierarchy is else consistent with~\figref{afl}.)}
\end{figure*}

Last, \figref{block} shows structural modules in \textit{javax} software and \textit{elegans} metabolic networks revealed with \cpa~and \hpa~algorithms. Observe that most prominent functional modules in \textit{javax} network are revealed in regions with lower clustering---indicated by strong off-diagonal structure---while communities mostly exist in regions with higher clustering. However, the latter is not the case for \textit{elegans} network, where all nodes exhibit very high clustering. Note that both structures revealed with \hpa~algorithm are still consistent with the structural-world conjecture, although different types of modules are better separated in the case \textit{javax} network (\secref{mix}). This is further indicated by higher clustering assortativity $r_d$, which equals $0.545$ for \textit{javax} network, while only $0.183$ for \textit{elegans} network. Nevertheless, structure of \textit{elegans} network might be modeled more adequately, when communities are extracted from the network apriori.


\section{\label{sec_cora}\cora}
In the following we conduct a more detailed analysis of a larger information network using techniques presented in the paper. We adopt a citation network extracted from the famous \textit{Cora} dataset~\cite{MNRS00} that includes computer science publications collected from the web, and publications automatically parsed from the bibliographies of the latter. Complete network reduced to the largest connected component contains $23166$ nodes and $89157$ links, while other common statistics are given in~\tblref{cora}.

\begin{table}[h]
\begin{ruledtabular}
\begin{tabular}{ccccccc}
Description & $n$ & $m$ & $k$ & $C$ & $r$ & $r_d$ \\\hline
\coras & $23166$ & $89157$ & $7.7$ & $0.266$ & $-0.055$ & $0.394$ \\
Community extraction & $14602$ & $29003$ & $4.0$ & $0.143$ & $-0.083$ & $0.547$ \\
Functional modules & $6832$ & $10345$ & $3.0$ & $0.073$ & $-0.143$ & $0.489$ \\
\end{tabular}
\end{ruledtabular}
\caption{\label{tbl_cora}Common statistics for \cora~before and after community extraction, and network induced by functional modules identified in the latter. (Networks are treated as simple undirected graphs.)}
\end{table}

\begin{figure*}[t]
\includegraphics[width=1.50\columnwidth]{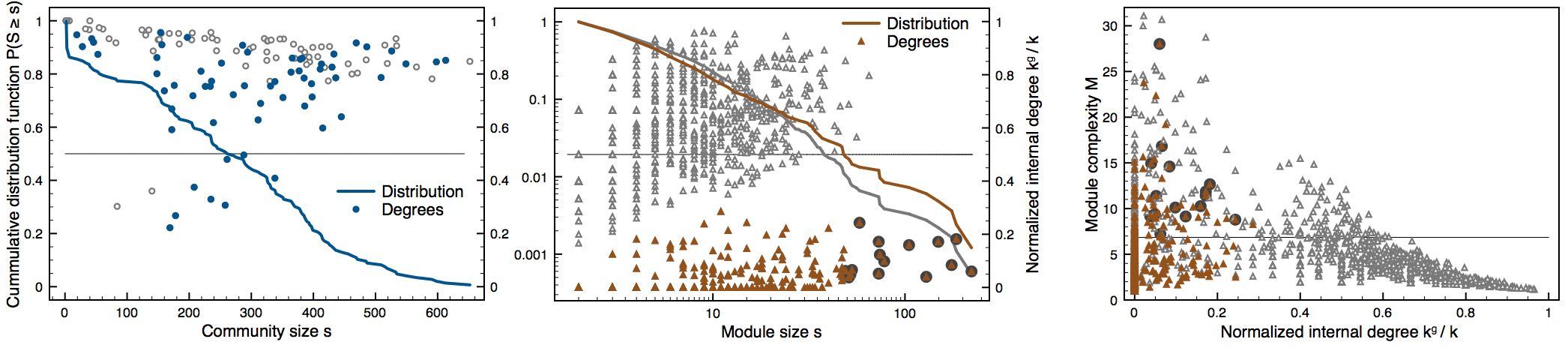}
\caption{\label{fig_coraM}(Color online)~Communities and functional modules within \cora\space (left-hand and right-hand side, respectively). (Filled symbols correspond to extracted communities and identified functional modules, where highlighted modules contain more than $48$ nodes. Horizontal line in the right-most plot represents average $M$ for functional modules.)}
\end{figure*}

We first employ \sff~and \lff~network models introduced in~\secref{mod} to analyze citation dynamics depicted in the network. \tblref{mod}~shows the networks generated by varying the burning and linking probabilities $p$ and $q$ within the models. Observe that, for $p\approx 0.3$ and $p\approx 0.75$, both models match common statistics of \coras~almost precisely. However, as already discussed in~\secref{mod}, clustering assortativity $r_d$ is underestimated.

\begin{table}[h]
\begin{ruledtabular}
\begin{tabular}{cccccccccc}
Model & $p$ & $q$ & $n$ & $m$ & $k$ & $C$ & $r$ & $r_d$ \\\hline

\multirow{5}{*}{\sff} & $0.275$ & $0.75$ & \multirow{5}{*}{$23166$} & $79886$ & $6.9$ & $0.263$ & $-0.060$ & $0.213$ \\
& \multirow{3}{*}{$0.3$} & $0.725$ & & $78724$ & $6.8$ & $0.270$ & $-0.056$ & $0.217$ \\
& & $0.75$ & & $\mathbf{87547}$ & $\mathbf{7.6}$ & $\mathbf{0.269}$ & $\mathbf{-0.057}$ & $\mathbf{0.192}$ \\
& & $0.775$ & & $97485$ & $8.4$ & $0.265$ & $-0.057$ & $0.179$ \\
 & $0.325$ & $0.75$ & & $96729$ & $8.4$ & $0.272$ & $-0.049$ & $0.179$ \\\hline

\multirow{5}{*}{\lff} & $0.25$ & $0.775$ & \multirow{5}{*}{$23166$} & $82057$ & $7.1$ & $0.265$ & $-0.059$ & $0.277$ \\
& \multirow{3}{*}{$0.275$} & $0.75$ & & $79581$ & $6.9$ & $0.275$ & $-0.051$ & $0.285$ \\
& & $0.775$ & & $\mathbf{88870}$ & $\mathbf{7.7}$ & $\mathbf{0.272}$ & $\mathbf{-0.053}$ & $\mathbf{0.273}$ \\
& & $0.8$ & & $100820$ & $8.7$ & $0.266$ & $-0.054$ & $0.270$ \\
 & $0.3$ & $0.775$ & & $97222$ & $8.4$ & $0.277$ & $-0.049$ & $0.274$ \\

\end{tabular}
\end{ruledtabular}
\caption{\label{tbl_mod}Statistics for networks containing $23166$ nodes generated with \sff~and \lff~network models. (Results are estimates of the mean over $10$ realizations of network models.)}
\end{table}

According to~\eqref{v}, the number of visited nodes $v$ by each newly added node within the models---the number of publications considered by the authors---can be estimated between $1.61$ and $1.75$ (simulations give similar results). As the average number of publications within each bibliography equals $k/2=3.84$, the fraction of publications considered by the authors, relative to the number of publications cited, is $2v/k\approx 0.45$. Hence, according to citation dynamics behind \coras, more than two times as many publications are cited than actually read.

Note that above results are somewhat influenced by (automatic) network sampling procedure (see~\cite{MNRS00} for details). The latter can be clearly observed in much lower average degree $k$ than in other citation networks (see, e.g., \textit{hepart} network in~\tblref{rw}). Furthermore, common density of real-world networks also estimates a much higher $k$~\cite{LJTBH11,BSB12}. It ought to be mentioned that scale-free exponents $\alpha$ for degree distributions of the networks generated with \sff~and \lff~models are $2.78$ and $2.67$, whereas \coras~exhibits $\alpha=3.3$. (Power-laws $p_k\propto k^{-\alpha}$ are plausible fits at $p\mbox{-value}=0.01$~\cite{CSN09}.)

We next also reveal different structural modules expressed in the structure of \coras. Identification proceeds as follows. First, dense structural modules are identified in the network based on the extraction framework presented in~\secref{mix}. The final pool consists of $146$ modules, whereas $61$ of these are extracted as communities according to a criteria based on a local maximum of $r_d$. Remaining network contains $14602$ nodes, while other common statistics are given in~\tblref{cora}. \figrefs{coraM}{left}~further shows different properties of identified modules.

Second, remaining structural modules in the network are revealed using the proposed \hpa~algorithm. The algorithm detects $1819$ groups of nodes, where $827$ of these are identified as functional modules according to~\figrefs{coraM}{middle}. Common statistics for the network induced by functional modules are given in~\tblref{cora}.

As predicted by the structural-world conjecture, identified communities and functional modules overlap considerably. $43\%$ of nodes in communities also appear in functional modules, whereas $41\%$ of nodes in functional modules are overlaid by communities. Each node is else in $1.25$ communities. \figref{coraM}~also shows distributions of module sizes $s$ (note different scales). Observe that distribution for communities is rather uniform (\figrefs{coraM}{left}), which is inconsistent with some earlier work~\cite{CNM04,PDFV05}. On the other hand, size distribution of functional modules shows a plausible fit to a power-law $p_s\propto s^{-\alpha}$ with $\alpha=2.24$ at $p\mbox{-value}=0.1$~\cite{CSN09} (\figrefs{coraM}{middle}).

\figrefs{coraM}{right}~plots the complexity $M$ for structural modules identified in \coras\space after extraction of communities. $M$ measures the complexity of linking between different structural modules. Let $p_g$ be the probability that a neighbor of a node in some module is in module $\g$ and let $G$ to be the corresponding random variable. Then, $M$ of concerned module is defined as
\begin{eqnarray}
M & = & b^{H(G)},
\label{eq_M}
\end{eqnarray}
where $H(G)$ is the entropy of $G$, $H(G)=-\sum_g p_g\log p_g$, and $b$ is the base of the logarithm, $M\geq 1$. Note that \eqref{M} is equivalent to Shannon's coding theorem~\cite{Sha48} (up to a constant). $M$~of a module is thus the expected number of dependent modules in the network. Hence, $M$ is close to one for communities and, e.g., functional modules that form bipartite structures, whereas higher values correspond to more complex configurations.

Values of $M$ shown in~\figrefs{coraM}{right} reveal than complexity of linking patterns between different modules is much higher than expected. For example, average $M$ for functional modules equals $6.85$, however, the result is influenced by a large number of smaller modules identified in the network (many small communities remain). Indeed, when the structure is reduced to $31$ modules that represent more than $48$ nodes in the original network, and only inter-dependencies that are supported by more than $48$ links are considered, average $M$ decreases to $1.84$. Nevertheless, functional modules still arrange in configurations that go beyond, e.g., simple bipartite structures.

\begin{figure*}[t]
\includegraphics[width=1.66\columnwidth]{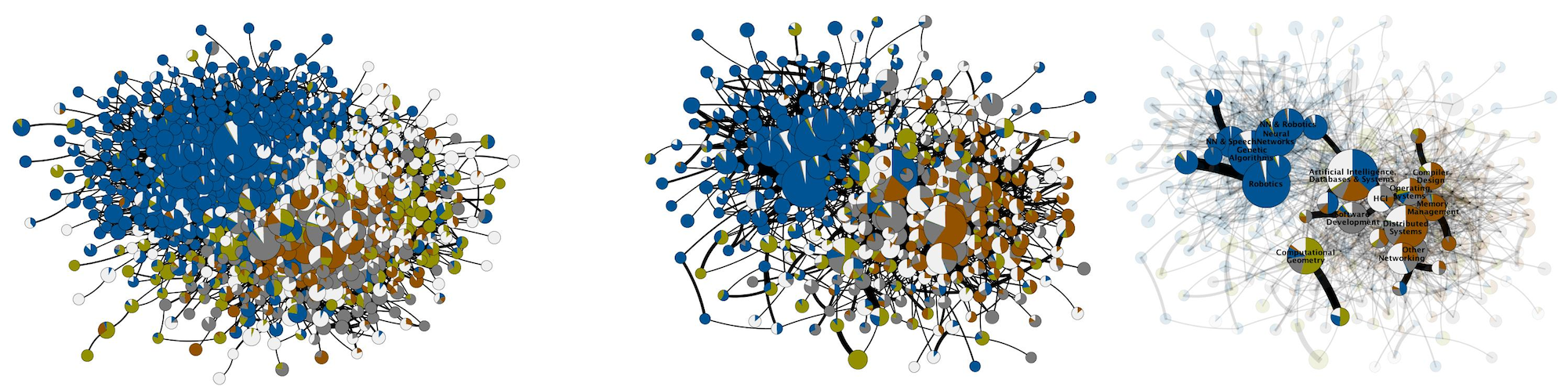}
\caption{\label{fig_coraO}(Color online)~Largest connected components of community and functional module overlays upon \cora~after extraction (left-hand and right-hand side, respectively). Modules in the right-most overlay correspond to more than $48$ nodes. (Node sizes are proportional to module sizes, whereas histograms represent high-level research topic classification in~\cite{MNRS00}---\textit{artificial intelligence} ({\color{blue}blue}), \textit{operating systems} ({\color{mocha}mocha}), \textit{data structures and algorithms} ({\color{asparagus}asparagus}), \textit{programming} ({\color{gray}gray}) and other---and are consistent among figures. Due to simplicity, modules representing four nodes or less are ignored.)}
\end{figure*}

Although not directly discussed above, general structural modules more accurately model \coras\space after extraction than a framework limited to communities. For example, average height of a hierarchy revealed with \hpa~algorithm equals $4.7$, whereas $7.0$ for \cpa~community detection algorithm (\secref{exp}). \figref{coraO}~also shows corresponding module overlays. Observe that functional modules recognize artificial intelligence and operating systems as two rather independent fields of computer science, which are related through interdisciplinary fields like data structures, algorithms and programming (\figrefs{coraO}{middle}). However, community structure of the network fails to acknowledge the latter (\figrefs{coraO}{left}).

\begin{table*}[t]
\begin{ruledtabular}
\begin{tabular}{p{1.45cm}ccl}
\multicolumn{1}{c}{Modules} & $s$ & $k$ ($n$) & \multicolumn{1}{c}{Reference / Topic} \\\hline

\multirow{4}{1.45cm}{Neural networks} & $20$ & $136$ & Hertz et al., \textit{Introduction to the theory of neural computation} (Addison-Wesley, 1991). \\\noalign{\smallskip}
& \multirow{3}{*}{$129$} & ($73$) & \textit{Artificial intelligence} -- \textit{machine learning} -- \textit{neural networks}. \\
& & ($11$) & \textit{Artificial intelligence} -- \textit{machine learning} -- \textit{probabilistic methods}. \\
& & ($6$) & \textit{Artificial intelligence} -- \textit{machine learning} -- \textit{genetic algorithms}. \\\hline

\multirow{4}{1.45cm}{Genetic algorithms} & $37$ & $182$ & Goldberg, \textit{Genetic algorithms in search, optimization and machine learning} (Addison-Wesley, 1989). \\\noalign{\smallskip}
& \multirow{3}{*}{$175$} & ($90$) & \textit{Artificial intelligence} -- \textit{machine learning} -- \textit{genetic algorithms}. \\
& & ($15$) & \textit{Artificial intelligence} -- \textit{machine learning} -- \textit{neural networks}. \\
& & ($12$) & \textit{Artificial intelligence} -- \textit{games and search}.  \\\hline

\multirow{6}{1.45cm}{Graph drawing} & \multirow{3}{*}{$9$} & $18$ & Kant, \textit{Drawing planar graphs using the lmc-ordering}, in \textit{Proc. of SFCS '92}, pp. 101-110. \\
 & & $18$ & Schnyder, \textit{Embedding planar graphs on the grid}, in \textit{Proc. of SODA '90}, pp. 138-148. \\
 & & $7$ & Kant et al., \textit{Area requirement of visibility representations of trees}, in \textit{Proc. of CCCG '93}, pp. 333-356. \\\noalign{\smallskip}

 & \multirow{3}{*}{$23$} & $11$ & Chrobak et al., \textit{Convex drawings of graphs in two and three dim.}, in \textit{Proc. of SCG '96}, pp. 319-328. \\
 & & $9$ & Garg et al., \textit{Planar upward tree drawings with optimal area}, Int. J. Comput. Geom. Ap., \textbf{6}, 333 (1996). \\
 & & $8$ & Garg, \textit{Where to draw the line}, PhD thesis, Brown University (1996). \\

\end{tabular}
\end{ruledtabular}
\caption{\label{tbl_pubs}Research topic classification~\cite{MNRS00} and publications corresponding to hub nodes within selected functional modules in~\figref{coraO}. Neural networks and genetic algorithms modules represent equally labeled bipartite structures, whereas graph drawing corresponds to lower-most pronounced bipartition (right-hand and left-hand side of~\figrefs{coraO}{right}, respectively). Results show hub nodes within smaller modules and topic distribution of larger modules (below and above, respectively), while only high degree nodes are reported for graph drawing. Note that highest degrees in the larger neural networks and genetic algorithms modules are else only $19$ and $26$. Also, interestingly, Roberto~Tammasia who was also a PhD advisor of Ashim~Garg has coauthored all papers on graph drawing with multiple authors. (References have been abbreviated to fit page width.)}
\end{table*}

Last, \tblref{pubs}~also shows research topic distributions~\cite{MNRS00} and publications representing hub nodes for selected functional modules in~\figref{coraO}. Similarly as in~\secref{exp}, the structure can be related to different roles of publications within a certain field. For example, largest functional modules represent publications addressing similar problems that depend on a smaller set of prior seminal publications or, e.g., book reviews. The latter commonly correspond to hubs in the network. Nevertheless, many of the revealed configurations of functional modules include no hub nodes. (See caption of~\figref{coraO}.)


\section{\label{sec_conc}Conclusions}
Findings in the paper expose functional modules as another key ingredient of complex real-world networks. These are together with communities combined into a structural-world conjecture, which provides a mesoscopic view on the structure of networks. We propose a natural model based on the latter that generates networks with most common properties, whereas we also introduce a simple algorithm that outperforms state-of-the-art in detection of structural modules. We further propose several other techniques, valid for exploratory network analysis. 

Future work will focus on the analysis of larger networks with millions of nodes and links, to devise a more detailed classification of structural-world networks.


\appendix


\begin{acknowledgments}
This work has been supported by Slovene Research Agency ARRS within Research Program No. P2-0359.
\end{acknowledgments}


\begin{figure*}[t]
\includegraphics[width=1.50\columnwidth]{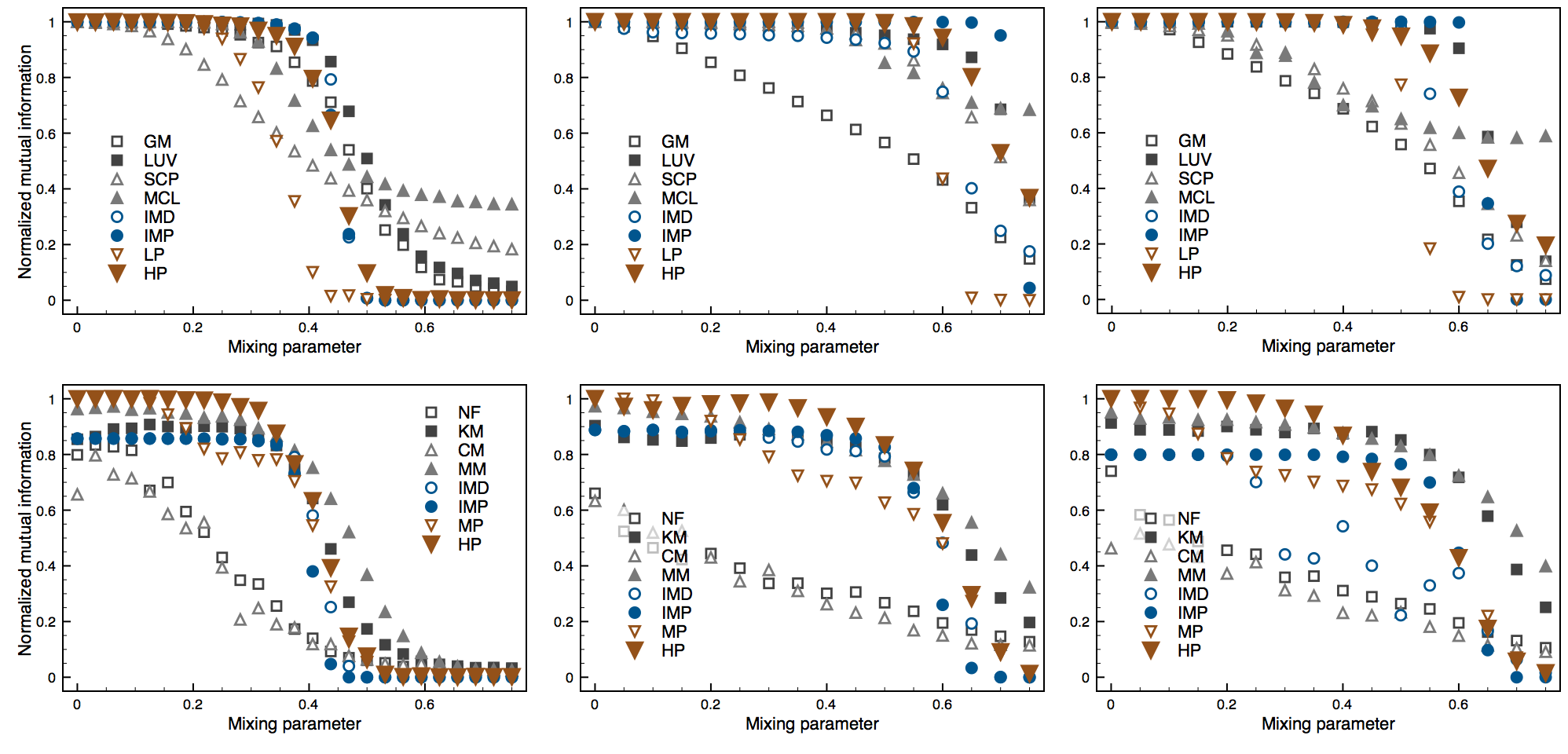}
\caption{\label{fig_soa}(Color online)~Comparison of module detection algorithms on~\GN~and~\LFR~synthetic networks with smaller and larger communities (top left to right, respectively), and on~\GNT, \HNV~and~\HNX~synthetic networks (bottom left to right, respectively). (Results are estimates of the mean over $100$ network realizations, and $10$ realizations for \mcl~and \cm~algorithms.)}
\end{figure*}

\section{\label{app_soa}State-of-the-art}
Proposed \hpa~algorithm is compared against different state-of-the-art algorithms for community detection, and for detection of structural modules. We adopt the following community detection approaches: greedy optimization of modularity~\cite{New04a,CNM04} (\gmo~algorithm), multi-stage optimization of modularity~\cite{BGLL08} (\luv~algorithm), sequential clique percolation method~\cite{KKKS08} (\scp~algorithm), Markov clustering algorithm~\cite{Don00} (\mcl~algorithm), structural compression method known as Infomod~\cite{RB07} (\imd~algorithm), random walk based compression known as Infomap~\cite{RB08} (\imp~algorithm), and a label propagation algorithm~\cite{RAK07} (\lpa~algorithm). (\scp~algorithm returns overlapping communities; thus, each node in multiple communities is classified into a random one.)

The algorithms are analyzed on three synthetic benchmark networks with planted communities. First is a classical community detection benchmark~\cite{GN02}, where the networks consist of four communities with $32$ nodes (\GN~networks). We also consider two variations of more realistic synthetic networks with scale-free degree and community size distributions~\cite{LFR08} (\LFR~networks). The number of nodes in the networks is set to $1000$, while community sizes vary between $10$ and $50$, and $20$ and $100$ nodes. For all networks, the structure is controlled by a mixing parameter $\mu$, $\mu\in[0,1]$. When $\mu$ equals zero, all links are placed according to the predefined partition, whereas for $\mu$ equal to one, the networks are completely random.

The results are shown in~\figrefs{soa}{top}. \hpa~algorithm is proven to be at least comparable to most of the approaches considered, yet it is outperformed by \luv~and \imp~algorithms. Note that the latter are among the best community detection algorithms in the literature, while their performance on these networks cannot be further improved~\cite{LF09b,For10}. On the other hand, there is also no guaranty that the construction of networks does not implicitly introduce functional modules (that are detected by \hpa~algorithm). Nonetheless, both \luv~and \imp~algorithms are still limited to communities and cannot detect other structural modules, which we address next.

We adopt the following general module detection algorithms: symmetric nonnegative matrix factorization~\cite{DLPP06} (\nmf~algorithm), $k$-means data clustering~\cite{Mac67} based on~\cite{LKC10} (\km~algorithm), mixture model using expectation-maximization~\cite{NL07} (\mm~algorithm), mixture model with degree corrections~\cite{KN11a} (\cm~algorithm), structural compression method~\cite{RB07} (\imd~algorithm), model-based propagation algorithm~\cite{SB11g} (\mpa~algorithm), and the best community detection algorithm considered above~\cite{RB08} (\imp~algorithm). (\nmf~and \cm~algorithms are applied to each network for ten times, while the best revealed partition is reported. \nmf, \km, \mm~and \cm~algorithms demand the number of modules apriori.)

The algorithms are compared on \GNT, \HNV\space and \HNX\space synthetic networks (\secref{exp}). All networks contain communities and functional modules, where the latter are connected into bipartite and tripartite structures. Again, the links in the networks are placed according to~$\mu$.

\figrefs{soa}{bottom}~shows the results of the comparison. Only \hpa~algorithm can accurately detect the modules planted into these networks, and \mpa~algorithm for small enough $\mu$, while most other approaches fail. \mm~mixture model also performs relatively well, whereas \cm~model that incorporates additional constraints is relatively unstable on these networks (due to increased complexity). Interestingly, simple \km~data clustering performs much better than many other state-of-the-art algorithms~\cite{LKC10}.

The analysis does not include approaches that group nodes based on their network properties~\cite{NM11b}. However, these also reveal groups that are not structural modules.


\section{\label{app_mod}Network modeling}
The section considers different network modeling techniques represented by module factors $\nu_g$ (\eqref{mod}). We compare the strategy adopted withing \hpa~algorithm with three alternatives, where $\nu_g$ equal to $0$, $0.5$ or $1$ for all modules $g$ (denoted \cpa, \dpa~and \fpa~algorithms). Hence, \cpa~algorithm is based on classical propagation between neighboring nodes that can detect merely communities. \fpa~algorithm propagates labels through common neighbors, which reveals functional modules. \dpa~algorithm is the default approach, where labels are simultaneously propagated between and through (common) neighbors.

The strategies are compared on~\GN~and~\GNT~synthetic benchmark networks (\appref{soa} and \secref{exp}). \GN~networks consist of only communities, while \GNT~networks contain communities and also functional modules. For both networks, the structure is controlled by a mixing parameter $\mu$, $\mu\in[0,1]$. For $\mu=0$, all links are placed according to the planted modules, whereas the networks are completely random for $\mu=1$.

\begin{figure}[t]
\includegraphics[width=1.00\columnwidth]{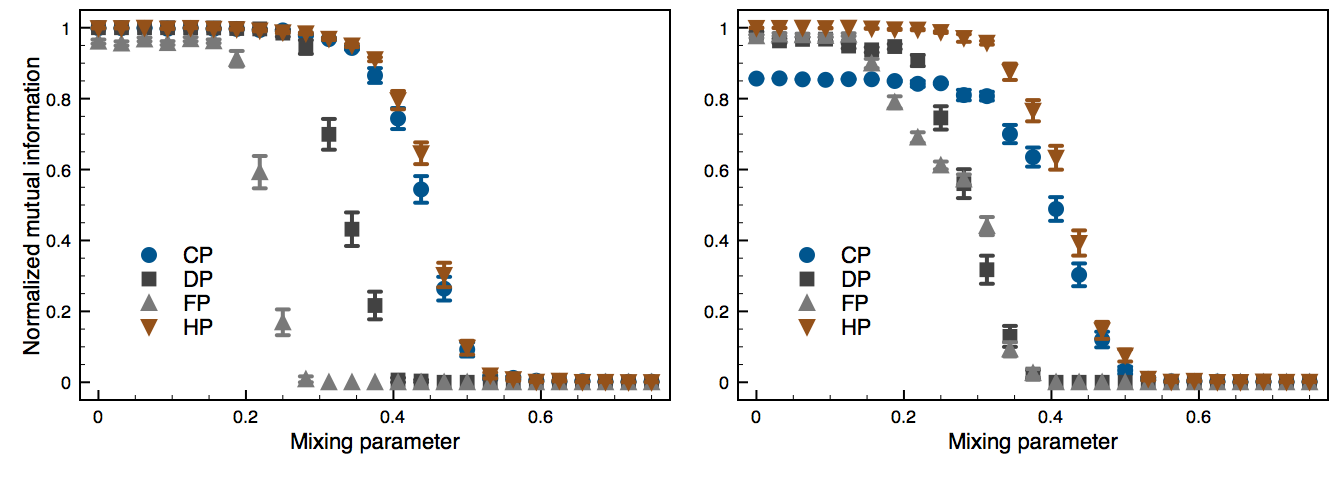}
\caption{\label{fig_mod}(Color online)~Comparison of network modeling on~\GN~and~\GNT~synthetic networks (left and right, respectively). (Results are estimates of the mean over $100$ network realizations, while bars show standard error of the mean.)}
\end{figure}

The results are shown in~\figref{mod}. \cpa~algorithm can accurately detect communities, while functional modules are identified as a single group of nodes. Interestingly, \fpa~algorithm can reveal functional modules and also communities, however, the stability of the algorithm is rather challenged. \dpa~algorithm performs reasonably well on both networks, although the results can be improved considerably (see \hpa~algorithm).

%


\end{document}